 [2005/06/22 5.2/AAS markup document class]%
\documentclass[preprint2]{aastex}%
\makeindex

\makeatletter 

\let\o@verbatim\verbatim
\def\verbatim{%
  \ifhmode\unskip\par\fi
  \ifx\@currsize\normalsize
     \small
  \fi
  \o@verbatim
}

\renewcommand \verbatim@font {%
  \normalfont \ttfamily
  \catcode`\<=\active
  \catcode`\>=\active
}

\RequirePackage{shortvrb}
\MakeShortVerb{\|}

\begingroup
  \catcode`\<=\active
  \catcode`\>=\active
  \gdef<{\@ifnextchar<\@lt\@meta}
  \gdef>{\@ifnextchar>\@gt\@gtr@err}
  \gdef\@meta#1>{\m{#1}}
  \gdef\@lt<{\char`\<}
  \gdef\@gt>{\char`\>}
\endgroup
\def\@gtr@err{%
   \ClassError{ltxguide}{%
      Isolated \protect>%
   }{%
      In this document class, \protect<...\protect>
      is used to indicate a parameter.\MessageBreak
      I've just found a \protect> on its own.
      Perhaps you meant to type \protect>\protect>?
   }%
}
\def\verbatim@nolig@list{\do\`\do\,\do\'\do\-}

\newcommand{\m}[1]{\mbox{$\langle$\it #1\/$\rangle$}}

\def\cmd#1{\cs{\expandafter\cmd@to@cs\string#1}}
\def\cmd@to@cs#1#2{\char\number`#2\relax}
\DeclareRobustCommand\cs[1]{\texttt{\char`\\#1}}
\def\GetFileInfo#1{%
  \def\filename{#1}%
  \def\@tempb##1 ##2 ##3\relax##4\relax{%
    \def\filedate{##1}%
    \def\fileversion{##2}%
    \def\fileinfo{##3}}%
  \edef\@tempa{\csname ver@#1\endcsname}%
  \expandafter\@tempb\@tempa\relax? ? \relax\relax}

\makeatother 

\title{Investigating slim disk solutions for HLX-1 in \\ ESO 243-49}

\author{O. Godet\altaffilmark{1,2}, B. Plazolles\altaffilmark{1,2}, 
  T. Kawaguchi\altaffilmark{3}, J.-P. Lasota\altaffilmark{4,5}, D. Barret\altaffilmark{1,2},
  S. A. Farrell\altaffilmark{6,7}, V. Braito\altaffilmark{6},
  M. Servillat\altaffilmark{8}, N. Webb\altaffilmark{1,2}, N. Gehrels\altaffilmark{9}}

\altaffiltext{1}{Universit\'e de Toulouse, UPS, Institut de Recherche en
  Astrophysique \& Plan\'etologie (IRAP), 9 Avenue du colonel
  Roche, 31028 Toulouse Cedex 4, France.}

\altaffiltext{2}{CNRS, UMR5277, 31028 Toulouse, France.}

\altaffiltext{3}{Center for Computational Sciences, University of Tsukuba,
1-1-1 Tennodai, Tsukuba, Ibaraki 305-8577, Japan}

\altaffiltext{4}{Institut d'Astrophysique de Paris, UMR 7095 CNRS, UPMC Univ Paris 06, 98bis Boulevard Arago, 75014 Paris, France.}

\altaffiltext{5}{Astronomical Observatory, Jagiellonian University, ul. Orla 171, 30-244 Kraków, Poland}

\altaffiltext{6}{Department of Physics and Astronomy, University of Leicester, University Road,
Leicester, LE1 7RH, UK.}

\altaffiltext{7}{Sydney Institute for Astronomy, School of Physics A29, The University of Sydney, NSW 2006,
Australia.}

\altaffiltext{8}{Harvard-Smithsonian Center for Astrophysics, 60 Garden Street, MS-67, Cambridge, MA 02138, USA.}

\altaffiltext{9}{NASA/Goddard Space Flight Center, Greenbelt, MD 20771, USA.}

\begin{abstract}

The hyper luminous X-ray source HLX-1 in the galaxy ESO 243-49, currently the
best intermediate mass black hole candidate, displays spectral transitions
similar to those observed in Galactic black hole binaries, but with a
luminosity 100-1000 times higher.  We investigated the X-ray properties of
this unique source fitting multi-epoch data collected by {\it Swift}, {\it
XMM-Newton} \& {\it Chandra} with a disk model computing spectra for a wide
range of sub- and super-Eddington accretion rates assuming a non-spinning
black hole and a face-on disk ($i=0^\circ$). Under these assumptions we find
that the black hole in HLX-1 is in the intermediate mass range ($\sim 2\times
10^4$ M$_\odot$) and the accretion flow is in the sub-Eddington regime. The
disk radiation efficiency is $\eta= 0.11\pm0.03$. We also show that the source
does follow the $L_X\propto T^4$ relation for our mass estimate. At the
outburst peaks, the source radiates near the Eddington limit.  The accretion
rate then stays constant around $4\times 10^{-4}$ M$_\odot$ yr$^{-1}$ for
several days and then decreases exponentially.  Such ``plateaus'' in the
accretion rate could be evidence that enhanced mass transfer rate is the
driving outburst mechanism in HLX-1. We also report on the new outburst
observed in August 2011 by the {\it Swift}-X-ray Telescope.  The time of this
new outburst further strengthens the $\sim 1$ year recurrence timescale.

\end{abstract}

\keywords{Galaxies: individual (ESO 243-49); Physical data and processes:
  Accretion, accretion disks, Black hole physics; Methods: data analysis; X-rays: individual (HLX-1)}

\expandafter\GetFileInfo\expandafter{\jobname.tex}%

\begin{document}

\maketitle
\section{Introduction}

Ultra Luminous X-ray sources (ULXs) are defined as off-nucleus extragalactic
sources showing X-ray luminosity exceeding $3\times 10^{39}$ erg s$^{-1}$
assuming isotropic emission (see Roberts et al. 2007). Even if their nature is
still in dispute, it is likely that their huge luminosity is produced by
accretion of matter onto a black hole (BH).  Three explanations for their
nature have been considered. (1) ULXs may be X-ray stellar-mass BH binaries
(BHBs) similar to those observed in our Galaxy, but in a more extreme version
of the very high state (e.g. Remillard \& McClintock 2006), the ultraluminous
or wind-dominated state (e.g. Gladstone et al. 2009). In such states, the
source would be able to radiate above the Eddington limit. It is still to be
understood why this ultraluminous state is so rarely observed in X-ray
binaries.  (2) King et al. (2001) proposed that the emission of ULXs is highly
anisotropic. In this case, the requirement to have super-Eddington emission is
alleviated. The nature of this anisotropy could be due to either geometrically
thick disks funnelling the X-ray photons produced in the inner parts of the
accretion disks (King 2009) or due to relativistic beaming of a jet or strong
outflow. However, the discovery of several optical and radio nebulae around
ULXs (e.g. Pakull \& Gris\'e 2008) and QPOs in M82 X-1 (Strohmayer \&
Mushotzky 2003) argue against strong beaming. (3) The final and the most
exciting explanation is that some ULXs are accreting intermediate mass BHs
(IMBHs) with masses ranging from $\sim 100~\mathrm{M}_{\odot}$ to
$\sim10^5~\mathrm{M}_{\odot}$ (e.g. Colbert \& Mushotzky 1999). The existence
of such IMBHs will naturally alleviate the need for super-Eddington
emission. It is clear now that the IMBH interpretation is not valid to explain
the ULX population as a whole (see e.g. Roberts 2007). However, the most
luminous ULXs, the so-called Hyper Luminous X-ray sources (HLX - Gao et
al. 2003) with X-ray luminosities above $10^{41}$ erg s$^{-1}$, are good
candidates. Finding convincing evidence for the existence of IMBHs is
important for the growth of supermassive BHs via mergers or accretion episodes
(e.g. Micic et al. 2007), dark matter studies (e.g. Fornasa \& Bertone 2008,
but see also Bringmann et al. 2009), cosmology (e.g. Trenti \& Stiavelli 2007)
and gravitational wave detection (e.g. Matsubayashi et al. 2004, Amaro-Seoane
\& Santamaría 2010).

Farrell et al. (2009) reported the serendipitous discovery of a ULX candidate
2XMM J011028.1--460421, referred to hereafter as HLX-1, located in the
outskirts of the edge-on spiral galaxy ESO 243-49 at a redshift of 0.0224
(Wiersema et al. 2010). From its maximum luminosity reaching $\sim 1.3\times
10^{42}$ erg s$^{-1}$ at peak and assuming that the source luminosity reached
up to ten times the Eddington limit ($L_{Edd}$), Farrell et al. (2009) derived
a BH mass of more than 500 M$_\odot$. HLX-1 is so far the best candidate to
harbour an IMBH. Apart from its extreme luminosity which has been observed
many times over the past 3 years by different X-ray satellites ({\it Swift},
{\it XMM-Newton} \& {\it Chandra}), HLX-1 is unique amongst other ULXs because
it is the only one for which clear spectral hysteresis similar to those
observed in Galactic BHBs (GBHBs) are seen. Indeed, Servillat et al. (2011)
using {\it XMM-Newton}, {\it Chandra} and {\it Swift} data confirmed the
results presented in Godet et al. (2009) that HLX-1 underwent transitions from
the high/soft state to the low/hard state. The {\it Swift}-X-ray Telescope
(XRT) lightcurve covers so far four outbursts from 2008 to 2011 (hereafter P0,
P1, P2 and P3 -- see Fig.~\ref{fig_LC}). While we had only one observation in
October 2008 during the P0 outburst, the P1 and P2 outbursts display
well-sampled Fast-Rise-Exponential-Decay (FRED)-like temporal profiles. Thanks
to our dedicated {\it Swift}-XRT ToO, we caught the rise of the P3 outburst
from 15$^\mathrm{th}$ August 2011 (Godet et al. 2011). A precursor could be
seen prior to the P3 outburst peak in Fig.~\ref{fig_LC}. Two reflare events
are also visible during the P1 and P3 outbursts. The outbursts appear to be
separated by a recurrence timescale of nearly a year. Recently, Lasota et
al. (2011) interpreted the X-ray light-curve as the result of enhanced
mass-transfer rate onto a pre-existing accretion disk around an IMBH, when an
asymptotic giant branch star orbiting along an eccentric orbit with a period
of $\sim 1$ year is tidally stripped near periastron.

In order to further investigate the nature of this unique source, it is
essential to put some constraints on the accretion flow and the BH mass. Lower
and upper limits from radio, optical and X-ray observations have been derived
(see Wiersema et al. 2011, Servillat et al. 2011 and Webb et al. 2011). Davis
et al. (2011) using their relativistic accretion disk model {\scriptsize
BHSPEC} to fit X-ray spectra when the source was in various spectral states
put some constraints on the BH mass. They showed that the inclination ($i$)
has a strong influence on both the mass and the spin ($a^*$) of the BH derived
by the model. Due to degeneracies in their best-fit parameters, they were only
able to give a range of possible BH masses from $3000~M_\odot$ (where the
limit corresponds to $i=0^\circ$, $\frac{L_X}{L_{Edd}}=0.7$ and $a^*=-1$) to
$3\times 10^5~M_\odot$ (where the limit corresponds to $i=90^\circ$ and
$a^*=0.99$). However, the inclination is likely to be less than $60-70^\circ$
due to the lack of observed eclipse dips assuming HLX-1 is a binary
system. These results were obtained neglecting the effects of radial advection
that are important at luminosities above the Eddington limit $L_{Edd}$ with
$L_{Edd} = 1.3 \times 10^{38}~\frac{M}{M_{\odot}}$ erg s$^{-1}$. Radial
advection is a mechanism enabled to stabilize disks in the super-Eddington
accretion regime (see e.g. Abramowicz et al. 1988).  This regime is reached
when the accretion rate ($\dot{M}$) is larger than $\dot{M}_{Edd} =
~\eta^{-1}\frac{L_{Edd}}{c^2}$ with $c$ the speed of light and $\eta$ the
radiation efficiency. In this regime, the vertical disk structure could be
geometrically thick ($H/R > 0.5$ with $H$ and $R$ the scale height and the
radius of the disk, respectively). However, for a range of moderate
super-Eddington $\dot{M}$-values there are still disk solutions for which the
vertical disk structure can be considered as relatively thin.  They are often
referred to as slim disk solutions (Abramowicz et al. 1988). For large
accretion rates ($\dot{m} \ge 10$ with $\dot{m}=\frac{\dot{M}~c^2}{L_{Edd}}$),
electron scattering (opacity and comptonization) have several effects on the
emergent spectra: i) they can be highly distorted and no longer look like a
multi-color black-body (BB) spectrum (e.g. Kawaguchi 2003, hereafter K03); ii)
the $L \propto T^4$ relation with $T$ the colour temperature at the inner disk
radius for standard $\alpha$-disk (Shakura \& Sunayev 1973) is no longer valid
(see Fukue 2000, K03). In addition, slim disks may extend beyond the innermost
stable circular orbit (ISCO) due to non-negligible pressure support resulting
in higher disk temperatures (e.g. Abramowicz et al. 1988, 2010; Watarai et
al. 2000).

In this paper, we make use of the K03 disk model in order to: i) investigate
the accretion flow and BH properties in HLX-1 through spectral fitting of a
multi-epoch and multi-instrument dataset using a model that computes a wide
range of sub- and super-Eddington accretion disk solutions; ii) investigate
whether radial advection and electron scattering have important effects on the
emergent spectra; iii) compare our results with those performed on less
luminous ULXs using the same model that always favoured critical or
super-Eddington accretion onto a stellar mass BH (e.g. Yoshida et al. 2010,
Foschini et al. 2006, Vierdayanti et al. 2006, Okajima et al. 2006).

The paper is organised as follows. In Section 2, we describe the data used to
perform the spectral analysis, as well as the data reduction. Section 3
presents some possible observational evidence for effects of radial advection
and/or electron scattering as well as the detection of possible X-ray
lines. In Section 4, we present the fitting results obtained with the K03 disk
model. In Section 5, we discuss the implications of the fitting results on the
BH mass, the accretion rate and the disk structure. The main conclusions are
given in Section 6.

\section{Observation log \& Data reduction}

We consider X-ray data coming from 3 observatories: {\it XMM-Newton}, {\it
  Chandra} (ACIS) and {\it Swift}. The data used in this study cover the
  different spectral states observed during the outbursts of HLX-1 (see
  Table~\ref{tab_obs}).  We used the same {\it XMM-Newton} and {\it Chandra}
  spectra as those presented in Servillat et al. (2011). The nomenclature used
  here is the same as that in Servillat et al. (2011) and Farrell et
  al. (2009) for the {\it XMM-Newton} data (see Table~\ref{tab_obs}).  For the
  {\it XMM-Newton} data, we only used the EPIC-pn data because it offers the
  best statistics for a given observation. Please refer to Servillat et
  al. (2011) for the details of the data reduction on the {\it XMM-Newton} and
  {\it Chandra} data. In order to investigate for the presence of X-ray lines
  during the outbursts, we also used the Reflection Grating Spectrometer (RGS)
  data from the XMM2 observation (see Table~\ref{tab_obs}).  The RGS data have
  been reduced using the standard {\it XMM-Newton} Science Analysis Software
  (SAS) task {\scriptsize RGSPROC} and the most recent calibration files
  released in February 2011. For the spectral analysis, we only considered the
  first order of the RGS data.

All the {\it Swift}-XRT Photon Counting data were processed using the HEASOFT
v6.11 and the lattest calibration files (CALDB version 3.8). This new CALDB
includes a new gain file enabled to correct the data for charge traps that
accumulate on the CCD due to radiation damage (see Pagani et al. 2011). These
traps can induce some energy offsets.  The data were processed using the tool
{\scriptsize XRTPIPELINE} v0.12.6. We inspected all segments of data to search
for epochs when the background was enhanced by the presence of hot pixels
induced by a high CCD temperature ($T > -55^\circ$). For these time intervals,
we checked whether there were some hot pixels in the extraction regions used
to extract the spectra.  Thus, we excluded the data from segments 00031287125
(0.8\,ks) and 00031287129 (0.2\,ks), because the data were severely
contaminated by hot pixels due to a high ($T > -51^\circ$C) CCD
temperature. We also excluded the first orbit ($\sim 0.6$ ks) of segment
00031287023 for the same reason. These latter bad data induce an artificially
high count rate in the X-ray light-curve at MJD $=55209$ (during the P1
outburst - see Fig.~\ref{fig_LC}). However, the re-flare event seen around
that time is not due to hot pixel contamination.  We used the grade 0-12
events, giving slightly higher effective area at higher energies than the
grade 0 events, and a 20 pixel (47.2 arcsec) radius circle to extract the
source and background spectra using {\scriptsize XSELECT} v2.4b. The
background extraction region was chosen in order to be close to the source
extraction region and not to contain any {\it XMM-Newton} sources.  The
ancillary response files were generated using {\scriptsize XRTMKARF} v0.5.9
and exposure maps generated by {\scriptsize XRTEXPOMAP} v0.2.7. The response
file {\scriptsize SWXPC0TO12S6$_{-}$20010101V012.RMF} is used to fit the
spectra.  To extract the {\it Swift}-XRT spectra, we divided the data over
different intensity ranges for the outbursts P0, P1 and P2. The work on the
first three outbursts being completed before the most recent outburst P3, we
decided to compare the latest outburst with the previous three by dividing the
data over time instead. Table~\ref{tab_nomenclature} summarises the
nomenclature used for the {\it Swift}-XRT data.

\section{Spectral analysis}
\label{spectral}

All the spectra were grouped at a minimum of 20 counts per bin to provide
sufficient statistics to use the $\chi^2$ minimisation technique.  The XMM3,
{\it Swift}-XRT and {\it Chandra} spectra were fitted in the 0.3-10 keV energy
range within {\scriptsize XSPEC} v12.7.0 (Arnaud 1996), while the XMM1 and
XMM2 spectra were fitted in the 0.2-10 keV energy range. The {\it Chandra}
data being moderately piled-up were fitted using the {\scriptsize PILEUP}
model (Davis 2001) within {\scriptsize XSPEC} with a frame time fixed at
0.8\,s (see Servillat et al. 2011 for more details). For each model, we
determined the appropriate value of the grade morphing parameter. For the data
used in this paper, the statistical errors are dominant over the instrumental
systematics. The RGS1 and RGS2 spectra were binned at twice the resolution of
the instrument ($\Delta \lambda= 0.2$ \AA). Since with this choice of binning
there are fewer than 20 counts per resolution bin, we used the C-statistic
(Cash 1979) available within {\scriptsize XSPEC} for the spectral fit.

The total column density was fixed at the best-constrained value of $N_H =
4\,\times\,10^{20}$ cm$^{-2}$ from the XMM2 observation (see Farrell et
al. 2009). The Galactic absorption column in the direction of the source is
equal to $1.8\,\times\,10^{20}$ cm$^{-2}$ (e.g. Kalberla et al. 2005). For all
fits, the absorption is modeled using the {\scriptsize WABS} photo-electric
absorption model. The source redshift being 0.0224 (Wiersema et al. 2010), we
adopted a source distance of $d_L=95$ Mpc using the cosmological parameters
from the WMAP5 results ($H_0 = 71$ km s$^{-1}$ Mpc$^{-1}$, $\Omega_M = 0.27$
\& $\Omega_\Lambda=0.73$). All the errors quoted below are given at a 90\%
confidence level for one parameter of interest ({\emph{i.e.}}
$\Delta\chi^2=2.706$).

\subsection{Hardness-Intensity diagram}
\label{HID}

We fitted the {\it Swift}-XRT spectrum using a multi-color disk {\scriptsize
DISKBB} and/or {\scriptsize POWERLAW} model. This phenomelogical model is
often used to fit the spectra of GBHBs. In that context, the powerlaw
component is often interpreted as emission produced by a corona of hot
electrons located in the inner regions of the accretion disk, while the
{\scriptsize DISKBB} component is interpreted as the emission from the
accretion disk (see Remillard \& McClintok 2006). The photon index of the
powerlaw was tied together for the $S_{CR0,1,2,3,4,5}$ spectra, because we
found $\Gamma$-values consistent within the errors when fitting the spectra
individually. The powerlaw normalisation and the parameters of the
{\scriptsize DISKBB} model were left free to vary independently between
spectra. The $S_{t1,2}$ spectra are well fitted by an absorbed {\scriptsize
DISKBB} model. We did not obtain a good fit ($\chi^2/dof=14.5/8$) using an
absorbed {\scriptsize DISKBB} model for the $S_{t3}$ spectrum. There are
residuals left at high energies. When adding a powerlaw component with
$\Gamma$ free, the $\Gamma$ parameter was not constrained. So, we decided to
freeze this parameter at the same value as found for the $S_{CR0,1,2,3,4,5}$
spectra ($\Gamma=2.2$). The fit is improved ($\Delta\chi^2=5$ for 1 dof
\emph{i.e.} an improvement at a $>2\,\sigma$ significance level). For each
spectrum, we then computed the hardness ratio (HR) of the 0.3-1.5 keV observed
flux over the 1.5-10 keV flux. This is different from the {\it Swift} hardness
ratio presented in Godet et al. (2009) and Servillat et al. (2011) that was
based on count rate. The HR values, the best-fit spectral parameters and the
0.2-10 keV unabsorbed luminosity ($L=4\pi ~d_L^2 ~F$ with $F$, the 0.2-10 keV
unabsorbed flux) for each spectrum are summarised in
Table~\ref{tab_diskbb}.

Fig.~\ref{fig_HID} shows the {\it Swift}-XRT hardness ratio versus the 0.2-10
keV unabsorbed luminosity. The points from the {\it Chandra} ($kT=0.21\pm0.01$
keV, $N=37^{+11}_{-8}$), XMM2 ($kT=0.17\pm0.01$ keV, $N=40^{+16}_{-10}$,
$\Gamma=2.1\pm0.4$, $N_\Gamma=1.9^{+0.8}_{-0.6}\times 10^{-5}$ ph cm$^{-2}$
s$^{-1}$ keV$^{-1}$) and XMM3\,\footnote{We also added a {\scriptsize MEKAL}
component with $kT=0.44^{+0.16}_{-0.12}$ keV to take a possible contamination
from the galaxy into account (see Servillat et al. 2011). Only the powerlaw
component was used to compute the hardness ratio.}
($\Gamma=2.2^{+0.4}_{-0.6}$ and $N_\Gamma=2.9\pm0.7\times 10^{-6}$ ph
cm$^{-2}$ s$^{-1}$ keV$^{-1}$) observations are also reported on the plot.

Fig.~\ref{fig_HID} shows that we have a good agreement between the different
instruments. The HLX-1 track in the Hardness-Intensity Diagram (HID) is
clearly reminiscent of that observed in GBHBs. Between $S_{1p+2p}$ and
$S_{after~2p}$, there is a softening of the source at nearly constant
luminosity ($\sim 1.2\times 10^{42}$ erg s$^{-1}$). Such a track in the HID is
observed in GBHBs (see e.g. Meyer-Hofmeister et al. 2009). However, the
situation appears to be different because the disk temperature near the peak
of the outbursts P1, P2 and P3 varies at nearly constant luminosity (from
$0.24\pm0.02$ keV to $0.18\pm 0.02$ keV -- see Table~\ref{tab_diskbb}). Note
that if we consider the $3\sigma$ errors on the temperature (\emph{i.e.}
$\Delta\chi^2 = 9$ for one parameter of interest), then we found that all data
have consistent $kT$-values given the limited statistical quality of the X-ray
data used. However, when merging the $S_{1p}$ \& $S_{2p}$ spectra
($S_{1p+2p}$) and the $S_{t1}$ \& $S_{t2}$ spectra ($S_{t1+t2}$), the
$kT$-values ($kT=0.240^{+0.023}_{-0.020}$ for $S_{1p+2p}$ and
$kT=0.20^{+0.019}_{-0.017}$ for $S_{t1+t2}$) are no longer consistent within
the $3\sigma$ errors.

\subsection{Investigating the $L-T$ relation}
\label{evidence}
\label{LT}

Fig.~\ref{fig_LxT} displays the disk temperature versus the unabsorbed 0.2-10
keV luminosity.  We fitted our $L-kT$ points using a $L\propto T^{a}$ relation
where $a$ was left as a free parameter. We obtained a good fit with $\log(L) =
(2.4 \pm 0.7)~\log~ T + 43.7 \pm 0.5$ ($1\,\sigma$ errors).  Feng \& Kaaret
(2007) showed that the luminosity $L$ varies as $L \propto T^{-3.1\pm0.5}$ in
the ULX NGC 1313 X-2. Using a larger sample of ULXs Kajava \& Poutanen (2008)
found a $L\propto T^{-4}$ relation. King (2009) suggested that the $L\propto
T^{-4}$ relation is expected in the super-Eddington regime provided that the
geometrical beaming factor varies as $b\propto \dot{m}^{-2}$. From
Fig.~\ref{fig_LxT}, the observed trend in the $L-kT$ space does not follow
such a correlation. Mizuno et al. (2001) found that some ULXs do present a
$L\propto T^2$ relation. They interpreted such a correlation as a possible
signature for moderate super-Eddington accretion (see also Fukue 2000;
K03). However, we cannot exclude a $T^4$ relationship at the $3\sigma$ level
(see also Servillat et al. 2011).  In Fig.~\ref{fig_LxT}, we also show the
  best-fit obtained using the $L\propto T^4$ relation.

\subsection{Evidence for possible line features}
\label{line}

When fitting the $S_{1p+2p}$ spectrum, we found some strong residuals around
0.6 keV (see the top left panel in Fig.~\ref{fig_line}). There may be some
lower significance residuals at higher energy as well. We separately
investigated the $S_{1p}$ and $S_{2p}$ spectra, and both show some residuals
around 0.6 keV in particular for $S_{2p}$. Given that the residuals are close
to the instrumental oxygen edge, we investigated if the residual may have an
instrumental origin. We extracted a background spectrum using different parts
of the field of view and encompassing the time when the source was at the
peak. The fit of the background spectrum does not reveal any residuals around
the O edge. We checked the images and found no evidence of contamination by
hot pixels. When the observations were performed the source location on the
CCD was not close to the bad columns. The application of the charge trap
correction did not change the residuals. So, the residuals are unlikely to be
instrumental. We added a Gaussian line redshifted at the galaxy distance
($z=0.0224$) with an intrinsic width equal to zero, because this parameter is
otherwise not constrained. The fit was improved by $\Delta\chi^2 = 12.7$ for 2
d.o.f. The line centroid is equal to $0.618^{+0.045}_{-0.042}$ keV and the
equivalent width is $EW=93^{+75}_{-51}$ eV (see the top right panel in
Fig.~\ref{fig_line}).  The disk parameters do not change with $kT =
0.25\pm0.01$ keV and $N=16^{+6}_{-4}$ (see Table~\ref{tab_diskbb} for a
comparison). In the rest of the paper, we did not include the Gaussian line
when fitting the $S_{1p+2p}$ spectrum, because it did not change the spectral
parameters of the K03 disk model considered in Section~\ref{ss} significantly.

In order to further investigate the soft X-ray emission at $\sim 0.6 $ keV, we
inspected the RGS data available from the XMM2 observation. The RGS 1 \& 2
spectra being background-dominated below 0.5 keV and above 1.3 keV, we only
considered this energy range in our analysis.  In the 0.5-1.3 keV energy band
a total of $1230$ and $1170$ counts were collected in the RGS1 and RGS2,
respectively (with 520 and 605 net counts). We first used an absorbed
{\scriptsize DISKBB} plus a power-law component to fit the data fixing the
photon index of the power-law component and the absorbing column density to
the values derived from the XMM2 data (Farrell et al. 2009, Servillat et
al. 2011).  We found a temperature ($kT=0.19\pm 0.02$ keV) consistent with
that derived from the EPIC data (see Table~\ref{tab_diskbb}; see also Farrell
et al. 2009, Servillat et al. 2011). Although the fit is statistically
acceptable (C-Statistic $= 160.4$ for 113 bins) a first inspection of the RGS1
\& 2 residuals (see the bottom panel in Fig.~\ref{fig_line}) reveals that
features are present at $\sim 0.6 $ keV, $\sim 0.9$ keV and $\sim 1$ keV.

We then added to the baseline continuum three unresolved Gaussian emission
lines redshifted at the galaxy distance.  The improvement in the fit due to
the addition of the $\sim 0.6$ keV emission line is $\Delta C/dof$ $= 26.2/2$
(corresponding to a detection significance $> 99.9$\% confidence). The
emission line is detected at an energy centroid of $E = 0.642 \pm 0.002$ keV,
a flux of $F = (1.91\pm 0.6) \times 10^{-5}$ photons cm$^{-2}$ s$^{-1}$ and an
equivalent width of $EW = 20 \pm 7$ eV. The energy centroid is consistent with
that derived from the {\it Swift}-XRT data, and the line equivalent width
seems to decrease with the luminosity of the continuum.  An additional line is
detected at $E=0.880_{-0.002}^{+0.008}$ keV ($\Delta C/dof$ $= 12.2/2$), with
a flux of $F=(0.93\pm 0.6) \times 10^{-5}$ photons cm$^{-2}$ s$^{-1}$
($EW=27\pm 12$ eV). Finally, the addition of the third Gaussian line at
$E=0.98\pm 0.01$keV ($EW=31\pm 22$ eV) gave a much lower statistical
significance ($\Delta C/dof$ $= 7.0/2$).  Even if additional data are needed
to further investigate the origin of these lines, we speculate that the most
likely identification for the $\sim 0.6$ keV and $\sim 0.9$ keV lines is
O{\scriptsize VIII} Ly$\alpha$ and Fe{\scriptsize XVIII}-Fe{\scriptsize
XIX}. The third line, if any, might be associated with Ne{\scriptsize X}
Ly$\alpha$.

\section{Investigating the accretion disk structure}
\label{ss}
\subsection{A simple approach}
\label{diskpbb}

In order to investigate whether the $L_X\propto T^{\sim 2.4}$ relation we
found in Section~\ref{evidence} could be the result of a slim disk surrounding
the BH, we first used the {\scriptsize DISKPBB} model instead of the
{\scriptsize DISKBB} model. The {\scriptsize DISKPBB} model is a
multi-temperature black-body disk model where local disk temperature is given
by $T(r)\propto r^{-p}$ with $p$, a free parameter (e.g. Mineshige et
al. 1994, Hirano et al. 1995, Watarai et al. 2000). When $p=0.75$, the model
is equivalent to a {\scriptsize DISKBB} model. For $p < 0.75$, radial
advection starts to become important (e.g.  Fukue 2000). Since the spectral
shape of the model could change significantly with the $p$-value, we fitted
the XMM1 and XMM2 spectra as well as those from {\it Swift} and {\it
Chandra}. The XMM3 spectrum was not considered because any evidence for a disk
component is only marginal (see Servillat et al. 2011). The results are
summarised in Table~\ref{tab_diskpbb}. In most cases, the $p$-values are not
well constrained. So, we cannot tell from this model whether or not radial
advection plays a role.

\subsection{The Kawaguchi (2003) disk model}
\label{slim}

\subsubsection{Description of the model}

Kawaguchi (2003) computed disk spectra in a self-consistent way, taking into
account the effects of electron scattering (opacity and disk Comptonization)
and the effects of the relativistic correction (\emph{i.e.}  gravitational
redshift and transverse Doppler shift) on the disk effective temperature in
the inner part of the accretion disk. Doppler boosting being not considered
implies that disk solutions are mostly seen face-on ($i=0^\circ$).  The model
table of the disk
spectra\footnote{http://heasarc.nasa.gov/xanadu/xspec/models/slimdisk.html},
from sub- to super-Eddington accretion rates with BH masses of $1 - 10^3
M_{\odot}$, was successfully used for analysis of several ULXs (e.g. Foschini
et al. 2006, Vierdayanti et al. 2006, Okajima et al. 2006, Yoshida et
al. 2010). We used here a new table including additional computation for
higher BH masses, extending up to $10^5 M_{\odot}$. The key parameters of the
model are the BH mass ($M$), the accretion rate ($\dot{M}$) and the viscosity
parameter ($\alpha$). For sub-Eddington accretion rates (\emph{i.e.} in the
case of the standard disk), the emergent disk spectra are
$\alpha$-insensitive. Near and at super-Eddington rates, the disk spectra
however become $\alpha$-sensitive, because electron scattering, which is
density sensitive (and density is $\alpha$-sensitive), begins to affect the
emergent spectra quite strongly. The normalisation is fixed using the source
distance ($d=95$ Mpc). To perform the fits, we considered the model option 7
that takes into account the effects of advection, electron scattering on
opacity, Comptonization and relativistic effects (see the
appendix~\ref{invest} for more details about the influence of the different
model options on the spectral parameters from HLX-1 data). It is important to
keep in mind that the K03 code automatically computes at each radius how much
advection is present for a given accretion rate. All the models are computed
assuming a non-rotating BH.

\subsubsection{Results}
\label{resultat}

To avoid calibration uncertainties between instruments affecting the fits
(see Tsujimoto et al. 2011), we decided to fit the spectra from
{\it XMM-Newton}, {\it Chandra} and {\it Swift} separately.

We first fitted the {\it Swift}-XRT spectra $S_{1p+2p}$, $S_{after~2p}$,
$S_{CR0}$ and $S_{CR3}$ by tying together the BH mass between them. We
proceeded in the same way with the viscosity parameter $\alpha$, while the
accretion rate was left as an independent parameter between the different
spectra. For the $S_{CR0}$ and $S_{CR3}$ spectra, we added a powerlaw
component of which the photon index was tied together between the two
spectra. This is because we found consistent values when fitting individually
the two spectra. When we fixed the $\alpha$-value to the default K03 model
($\alpha=0.01$) we did not obtain a good fit because some spectra display
strong residuals. In order to reproduce the spectra, it was necessary to leave
the viscosity parameter $\alpha$ free. We then obtained a good fit
($\chi^2/d.o.f. = 108/104$). The best-fit parameters are given in
Table~\ref{tab_slimdisk}. The derived BH mass is $M = 1.8^{+0.2}_{-0.1} \times
10^4~M_{\odot}$. Fitting the data from the outburst P3 using the K03 disk
model, we found $M=1.9^{+1.9}_{-0.2}\times 10^4$ M$_\odot$.  These estimates
are consistent and well inside the IMBH mass range. We note that the powerlaw
component was not needed to obtain a good fit for the $S_{t3}$ spectrum when
leaving the viscosity parameter free to vary.  Leaving the $N_H$-value free to
investigate the sensitivity of the BH mass with respect to the $N_H$ value did
not change the spectral parameters much and they are still consistent within the
errors with the values derived with $N_H=4\times 10^{20}$ cm$^{-2}$. Indeed,
we found $N_H=6.5^{+3.8}_{-2.5}\times 10^{20}$ cm$^{-2}$ and
$M=2.8^{+1.7}_{-0.9}\times 10^4$ M$_\odot$ from the outbursts P0-P2, while we
found $N_H= 4.1^{+2.8}_{-2.7} \times 10^{20}$ cm$^{-2}$ and
$M=1.9^{+3.2}_{-0.6}\times 10^4$ M$_\odot$ from the outburst P3.

Second, we fitted the {\it Chandra} spectrum. For the K03 disk model, we
estimated for the {\scriptsize PILEUP} model a grade morphing parameter of
0.33. Note that this parameter does not have a strong impact on the spectral
parameters derived here. The derived BH mass is $M = 1.9\pm0.2\times
10^4~M_{\odot}$. We then left the $N_H$-value free and we found
$N_H=3.7^{+3.9}_{-2.5}\times 10^{20}$ cm$^{-2}$ and a BH mass estimate ($M =
1.8^{+1.0}_{-0.4}\times 10^4$ M$_\odot$) consistent with those derived from
{\it Swift}.

Even if Farrell et al. (2009) showed that the XMM1 spectrum is well fitted by
a steep powerlaw ($\Gamma\sim 3.4$) suggesting that the source was in the
steep-powerlaw state as seen in some GBHBs, we decided to fit the XMM1 spectrum
as well as the XMM2 one with the K03 disk model. This is because the emergent
spectrum from the K03 disk model can have a shape strongly different from that
of a simple MCD spectrum (see Appendix~\ref{invest}). We tied the BH mass
between the two spectra, but we left $\alpha$ and $\dot{M}$ to vary
independently. First, we keep $N_H$ fixed at $4\times 10^{20}$ cm$^{-2}$. For
XMM2, it was necessary to add a powerlaw component to obtain a good fit, while
the addition of a powerlaw for XMM1 did not improve the fit.  The derived BH
mass is then $M = 1.4\pm0.1 \times 10^4$ M$_{\odot}$; which is not consistent
with the {\it Swift}-XRT and {\it Chandra} estimates within the 90\%
errors. If we consider the $3\sigma$ errors, then the estimates from the
different instruments are all consistent within the errors. Then, in order to
investigate the sensitivity of the BH mass with respect to the $N_H$ value, we
left free the $N_H$ parameter. We obtained a value of
$N_H=5.5^{+1.1}_{-1.0}\times 10^{20}$ cm$^{-2}$ and a BH mass of $M =
1.9^{+1.3}_{-0.3}\times 10^4~M_{\odot}$. The photon index of the powerlaw
component in both cases is consistent within the 90\% errors between the {\it
XMM-Newton} and {\it Swift}-XRT spectra. The best-fit results are given in
Table~\ref{tab_slimdisk}.

For XMM1, the accretion rate ($\dot{m}\sim 4.4$) and the luminosity that is
very different from that derived using a powerlaw ($L\sim 1.3\times 10^{42}$
erg cm$^{-2}$, a luminosity which was observed at the peak of the three
outbursts) are commensurate with the values found for the XMM2 and $S_{CR0}$
spectra (see Table~\ref{tab_slimdisk}). This however implies a large value of
$\alpha = 0.13\pm0.07$, even if all $\alpha$-values are consistent within the
$3\sigma$ errors.  If we force the viscosity parameter to the lowest possible
value of the K03 model ($\alpha=0.01$) keeping the mass fixed at $M=1.8~\times
10^4$~M$_\odot$ and $N_H$ free, we did not obtain a good fit
($\chi^2/dof=47.3/24$). The addition of a powerlaw improves the fit
($\chi^2/dof=23.3/22$), but in this case the powerlaw component dominates over
the disk component. In this case, we found similar $\Gamma$ and $N_H$ values
to those given in Farrell et al. (2009).

\section{Discussion}

\subsection{Constraints on the BH mass}
\label{unique}

To be able to derive a dynamical measurement of the BH mass in HLX-1 will be
very challenging given the distance of the source. So, we have to rely on
indirect estimates such as spectral fitting of the X-ray data. Zampieri \&
Roberts (2009) showed that the BH mass estimate derived using spectral fitting
can be highly variable depending on the disk model used (see their Table
2). In the case of HLX-1, Davis et al. (2011) used their advanced relativistic
disk model to fit spectra with different shapes. They found a BH mass within
the IMBH range (from 3000~M$_\odot$ to $3\times 10^5$ M$_\odot$) with extreme
and opposite assumptions. In this paper, we used the K03 disk model to fit
X-ray spectra of HLX-1 when the source was in various spectral states.  We
note that this model includes spectra for a wide range of accretion rates from
sub- to super-Eddington rates. In all cases for ULXs, the fits favoured super-
or near-Eddington accretion onto a stellar mass BH. However, the fits to HLX-1
spectra do favour an IMBH solution with a BH mass of
$M=1.8^{+1.6}_{-0.5}\times 10^4~M_\odot$. This is an interesting result
because we get three independent and nonetheless consistent mass estimates
from three different instruments ({\it Swift}-XRT, {\it XMM-Newton} EPIC-pn
and {\it Chandra}/ACIS) and over different spectral shapes. This estimate is
consistent with the observational lower and upper limits on the BH mass
(Farrell et al. 2009, Servillat et al. 2011).  So, using the K03 model which
assumes a non-spinning BH and a face-on accretion disk, we find that an IMBH
solution is favoured with an estimated mass of $\sim 1.8\times 10^4~M_\odot$.
The BH accretes at the Eddington limit ($\dot{m}\sim 10$) and radiates close
to the Eddington luminosity at the outburst peak ($L_{Edd} =
2.3^{+2.0}_{-0.6}\times 10^{42}$ erg s$^{-1}$). This corresponds to an
Eddington fraction of $f_{Edd} = 1.1^{+0.6}_{-0.5}$ considering a bolometric
luminosity of $2.5\times 10^{42}$ erg s$^{-1}$ on average (see
Fig.~\ref{fig_L_Mdot}).

\subsection{Comparison with the results from Davis et al. (2011)}


Using their BHSPEC model, Davis et al. (2011) studied the dependency of the BH
mass in HLX-1 with the inclination and the BH spin. We can only compare our
results with their work for $a^*=0$ and $i=0^\circ$. In that configuration,
they found a consistent mass estimate for XMM2 and {\it Chandra} with $\log M
\sim 3.8 \pm 0.1$ (Davis 2011, private communication). However, no
solution could be found in this case for the {\it Swift} spectra, because the
$f_{Edd}=L_X/L_{Edd}$ ratio was pegged to the maximum permitted value in their
BHSPEC model (\emph{i.e.}  $f_{Edd}=1$, Davis 2011, private
communication).  From Figures 2, 5 and 6 in Davis et al. (2011), the best-fit
contours for fixed inclination in the $a^*$ versus $\log M$ plots overlap well
between the simulations for the range of allowed parameters. Thus, they found
a consistent mass estimate between the three instruments for $i=0^\circ$ \&
$a^*=0.7$ with $\log M \sim 4.0$ for {\it Swift} \& {\it Chandra} and $\log M
\sim 3.9$ for XMM2. The mass values derived by Davis et al. (2011) from
the XMM2 and {\it Chandra} spectra for $a^*=0$ \& $i=0^\circ$ are smaller than
the one derived using the K03 disk model. We note that the Eddington ratios
($f_{Edd}$) they obtained are large (and close to 1). From our fitting results
using the K03 model, we found similar high $f_{Edd}$ values near the outburst
peak. We believe that for such high $f_{Edd}$ values the effects of advection
on the emergent spectra should be considered.

Evaluating the level of systematic errors of the K03 model by comparison with
alternative models by exploring the whole parameter space is beyond the scope
of the present paper and deserves a dedicated paper.  Nevertheless, we
estimated the level of systematic errors of the K03 model by comparison with
the KERRBB (Li et al. 2005) and BHSPEC models assuming a non-spinning BH
($a^*=0$) and a face-on disk ($i=0^\circ$) in the limit of low mass accretion
rate (\emph{i.e.} a few percent of the Eddington limit). In that limit, the
accretion disk is expected to be very close to the relativistic model of
Novikov \& Thorne (1973). We found that the level of systematic in the BH mass
between the K03 and KERRBB models is less than 45\% when only considering the
relativistic effects in both models (see Appendix~\ref{invest2}). When
comparing the results between the K03 and BHSPEC models taking into account
the relativistic effects and the effects due to Comptonization and electron
scattering, we found that the level of systematic errors in the BH mass is
less than $24\%$ (see Appendix~\ref{invest2}). In all cases, the values
of the accretion rate derived using the K03, KERRBB and BHSPEC models agree
within the errors. So, providing similar level of systematics errors over the
whole parameter space, we are confident that our BH mass and accretion rate
estimates are reliable.


\subsection{The disk structure}
\label{structure}

Fig.~\ref{fig_L_Mdot} shows a linear correlation between the 0.01-20 keV
bolometric disk luminosity ($L$) and the accretion rate so that
$L=\eta~\dot{m}~L_{Edd}$ with radiation efficiency $\eta =
0.11\pm0.03$. Watarai et al. (2000) found a similar linear correlation for
$\dot{m} < 20$. This implies that advection does not play a significant role,
and that the slim disk solutions are not needed. At all times during the
outbursts, the disk aspect ratio $H/R$ is less than 0.2, from Fig.~5 in
K03. Note that $H/R$ is mostly independent from the BH mass and the viscosity
parameter.

From Fig.~1 in K03, the disk appears to be radiation-pressure dominated in its
inner regions for all the $\dot{m}$-values found. Such disks might be
viscously and thermally unstable, and they might give rise to outbursts (Taam
\& Lin 1984; Honma et al. 1991, Lasota \& Pelat 1991; Xue et
al. 2011). However, Hirose et al. (2009) showed via MHD simulations that such
disks could be thermally stable (but see Xue et al. 2011). Most observations
of Galactic X-ray binaries in outburst radiating above $10\%$ of the Eddington
luminosity for a stellar mass BH do not show any evidence for such an
instability except maybe for GRS 1915+105 (Belloni et al. 1997; Xue et
al. 2011).

\subsection{The evolution of the accretion rate}

\begin{figure*}
\begin{tabular}{cc}
\hspace{-2cm}\includegraphics[angle=0,scale=0.55]{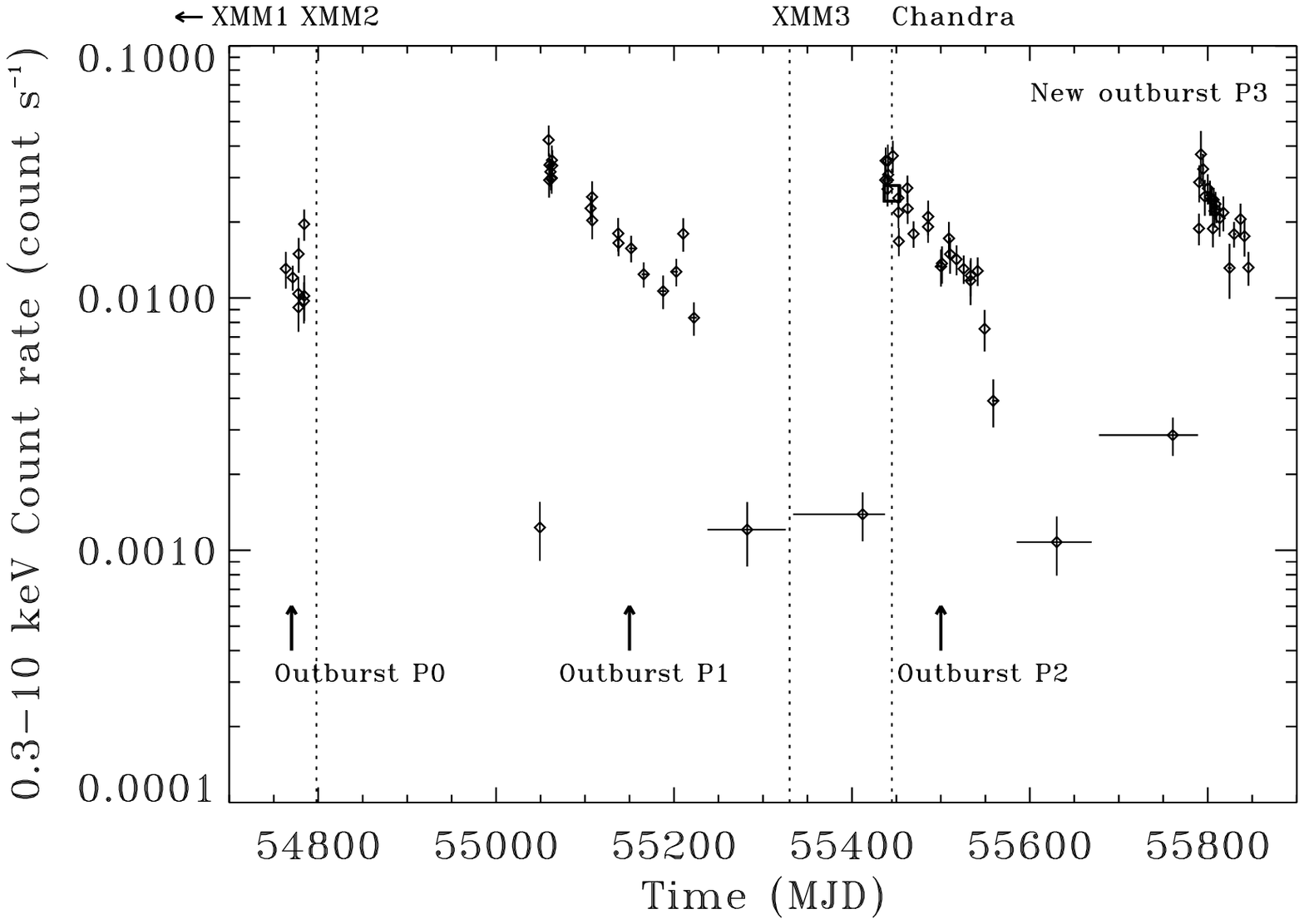} &
\includegraphics[angle=0,scale=0.52]{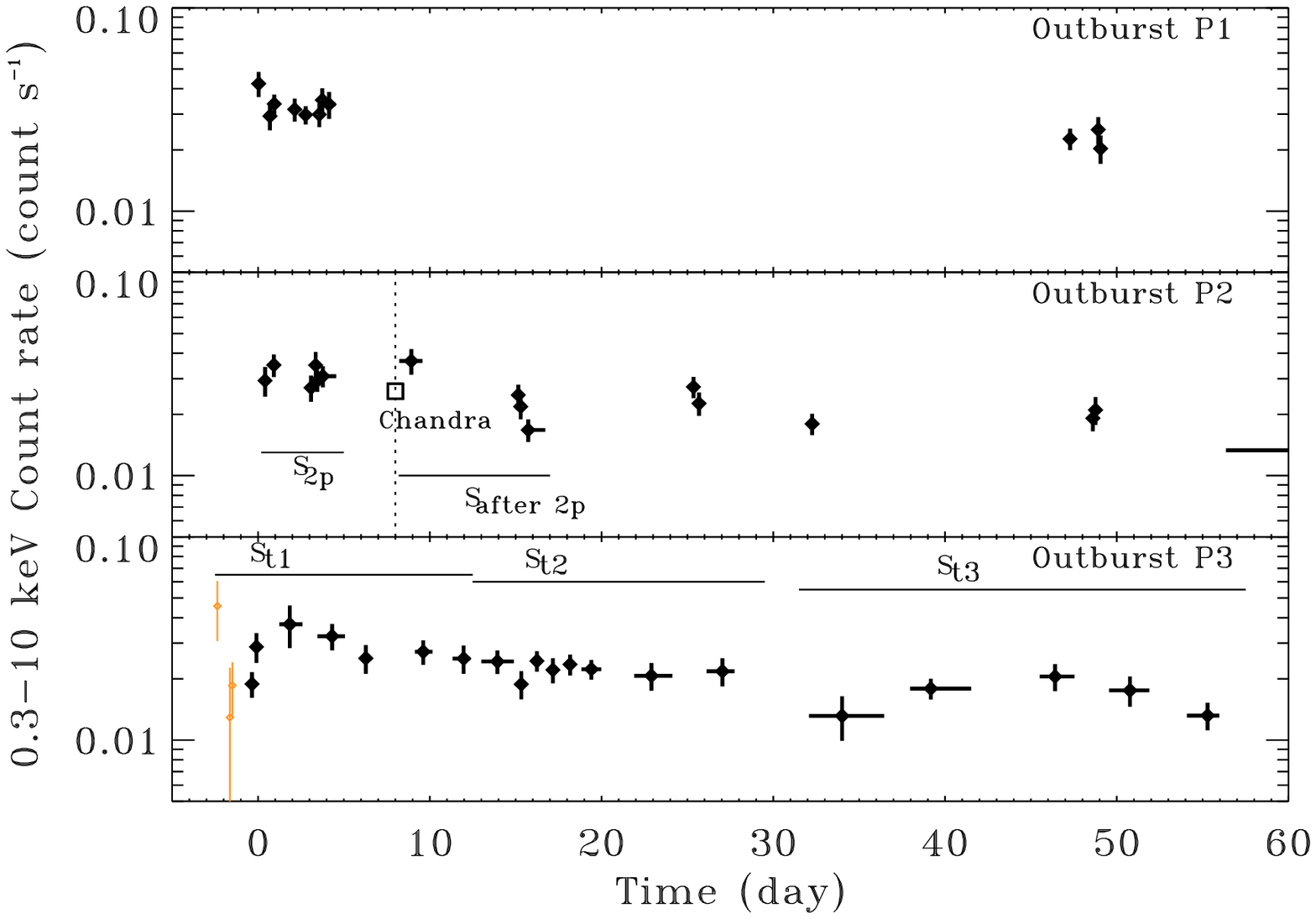} \\
\end{tabular}
\caption{{\it Swift}-XRT Photon Counting 0.3-10 keV light-curve of HLX-1 up to
  2011-09-05. The {\it Swift}-XRT light-curve was obtained using the {\it
  Swift}-XRT light-curve generator web interface with a binning of at least 60
  counts per bin (Evans et al. 2009). {\it Left --} The vertical dotted lines
  mark the times of the {\it XMM-Newton} and {\it Chandra} observations. {\it
  Right --} close-up of the X-ray light-curve around the outburst peaks P1, P2
  and P3. The zero time corresponds to the time when we first observed a
  re-brightening from the source (P1 $=$ 2009-08-16 and P2 $=$ 2010-08-29) and
  the peak of the outburst P3 (2011-08-15). The square on the middle panel is
  the predicted on-axis {\it Chandra} count rate into the 0.3-10 keV {\it
  Swift}-XRT energy band. Bottom panel: the thin points were obtained using a
  lower temporal binning for data from MJD $= 55788$ to MJD $=55790$. }
\label{fig_LC}
\end{figure*}

\begin{figure*}
\hspace{2cm}\includegraphics[angle=0,scale=0.65]{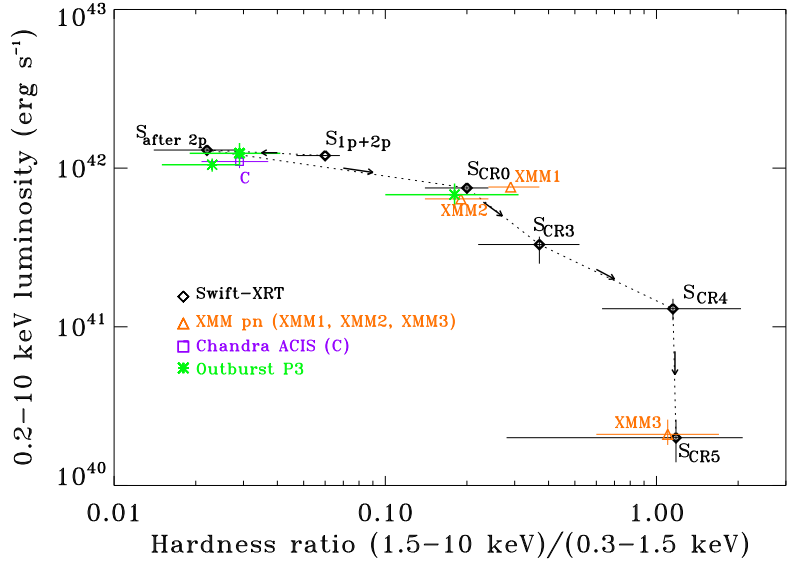}
\caption{{\it Swift}-XRT hardness versus intensity diagram. The {\it
    XMM-Newton} (triangles) and {\it Chandra} (square) points are also
    reported on the diagram. The arrows indicate the evolution from the peak
    to the end of the outbursts. }
\label{fig_HID}
\end{figure*}

\begin{figure*}
\includegraphics[angle=0,scale=0.55]{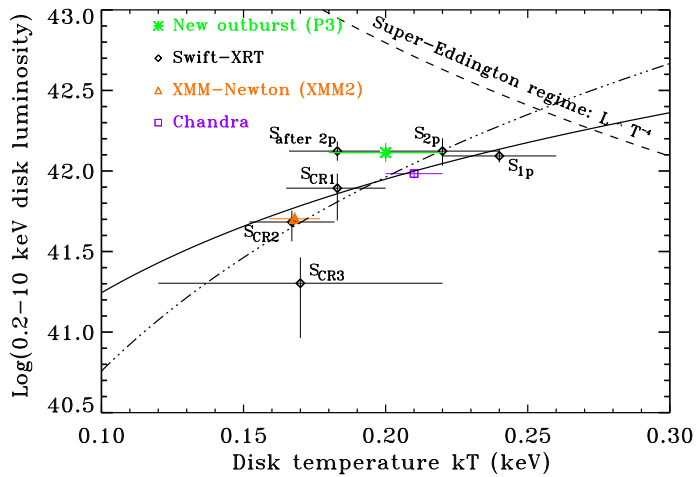}
\caption{Evolution of the unabsorbed 0.01-20 keV disk luminosity with the disk
  temperature. The values for {\it Swift} (diamonds), {\it XMM-Newton} (XMM2:
  triangles in orange) and {\it Chandra} (square in blue) are shown on the
  plot. The thick dotted-dashed line corresponds to the best fit obtained using the
  $L\propto T^4$ relation, while the thick solid line corresponds
  to the best fit obtained using the $L\propto T^\alpha$ relation with
  $\alpha\sim 2.4$.  The thick dashed line corresponds to the $L\propto
  T^{-4}$ prediction in the super-Eddington regime (Fukue 2000; King 2009). }
\label{fig_LxT}
\end{figure*}

\begin{figure*}
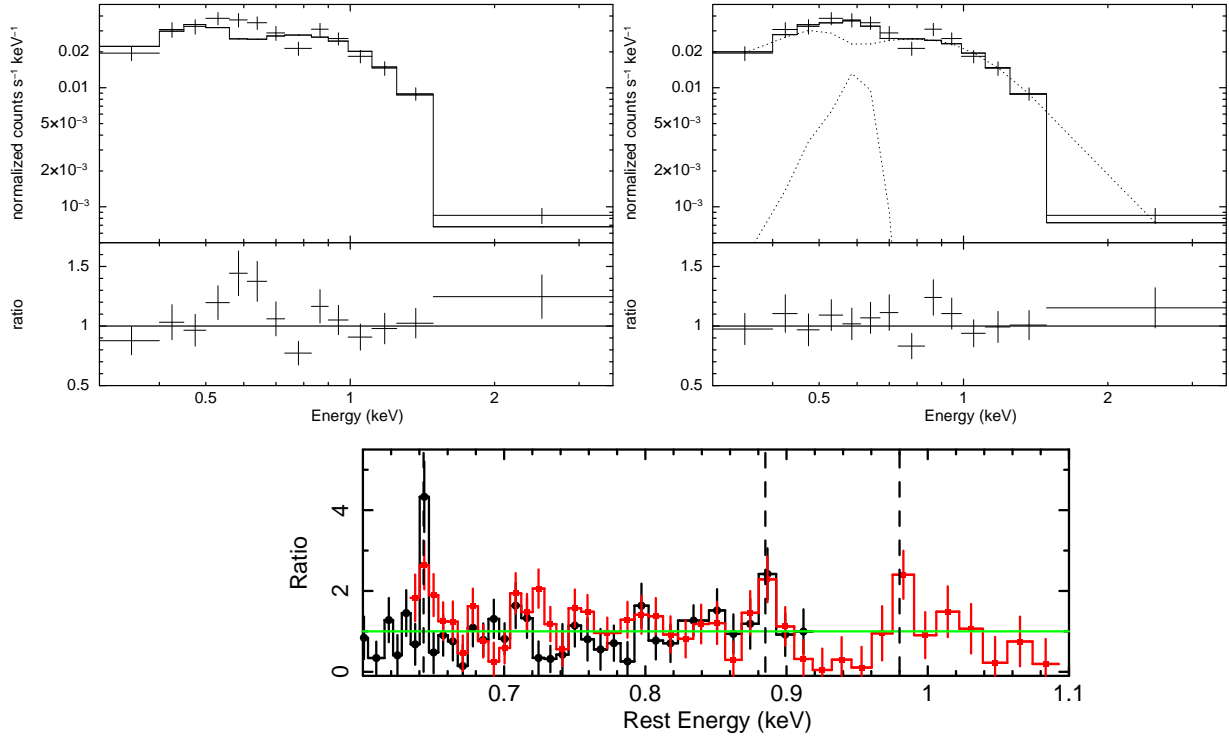

\begin{tabular}{c}
\includegraphics[angle=-90,scale=0.32]{plot_spectrum_pic1-2_diskbb_res.ps} 
\includegraphics[angle=-90,scale=0.32]{plot_spectrum_pic1-2_diskbb_zgauss.ps} \\

\hspace{2cm}\includegraphics[angle=-90,scale=0.4]{new.ps} \\
\end{tabular}
\caption{X-ray lines present in the {\it Swift}-XRT and {\it XMM-Newton} RGS
  data. {\it Top --} {\it Swift}-XRT spectrum when the source luminosity was
  at peak for the first and second outbursts ($S_{1p+2p}$). {\it Left:} using
  only an absorbed {\scriptsize DISKBB} model. {\it Right:} Same model with
  the addition of a Gaussian line redshifted at the galaxy redshift ($z =
  0.0224$). The dotted lines correspond to the different components of the
  model. {\it Bottom --} ratio of the RGS 1 (red) and RGS2 (black) spectra from
  the XMM2 observation over the model (an absorbed {\scriptsize DISKBB $+$}
  powerlaw model). The vertical dashed lines correspond to the centroids of the
  detected emission lines.}
\label{fig_line}
\end{figure*}

\begin{figure}
\includegraphics[angle=0,scale=0.5]{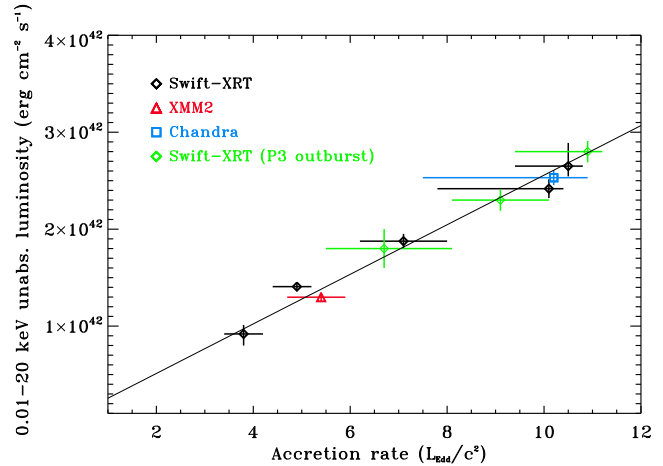}
\caption{Linear correlation between the 0.01-20 keV unabsorbed luminosity
  derived using the Kawaguchi (2003) slim disk model and the accretion
  rate. The luminosity was computed using the following relation:
  $L_{bol}=4\pi~d_L^2~F_{bol}$ with $F_{bol}$, the bolometric unabsorbed
  flux and $d_L=95$ Mpc.}
\label{fig_L_Mdot}
\end{figure}

\begin{figure}
\includegraphics[angle=0,scale=0.5]{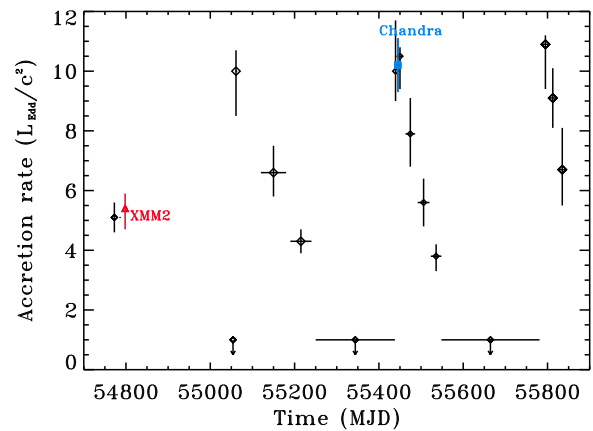}
\caption{Evolution of the accretion rate over time. }
\label{fig_Mdot_time}
\end{figure}

Fig.~\ref{fig_Mdot_time} shows the evolution of the accretion rate through the
different outbursts.  At the peak of the three outbursts and for a few weeks
($\sim 3$ and $\sim 2-3$ weeks for the outbursts P2 and P3, respectively), the
BH accretes at the Eddington limit ($\dot{m}\sim 10$). This ``plateau'' is
seen in the X-ray light-curve of the outbursts P2 and P3 (see the right panel
in Fig.~\ref{fig_LC}). Recently, Lasota et al. (2011) proposed that the HLX-1
outbursts may be due to enhanced mass-transfer rate onto a pre-existing
accretion disk when an asymptotic giant branch star orbiting along an
eccentric orbit with a period of $\sim 1$ year passing at periastron is
tidally stripped. Such a mechanism is known to produce ``plateaus'' in
light-curves (Bath \& Pringle 1981; Esin, Lasota \& Hynes 2000). The latter
authors considered an unstable disk that became stable after an enhanced
mass-transfer rate phase. In the case of HLX-1, the situation is different
since the disk is stable. Therefore, quasi-constant enhanced
mass-transfer rate during a given time interval would naturally produce a
``plateau'' in the light-curve.  After the ``plateau'', the accretion rate
drops exponentially with a decay time of $182\pm30$ days and $92\pm15$ days
for the outbursts P1 and P2, respectively. The difference in the decay time is
because there was a small re-flare event at the end of the outburst P1 (see
Fig.~\ref{fig_LC}). From the evolution of the accretion rate over time, we
calculated the mass accreted during the outbursts P1 and P2 assuming that the
accretion rate is the same throughout the disk. If it is not the case, then
the values derived below are lower limits. We found an accreted mass of $\sim
1.2\times 10^{-4}~M_\odot$ and $\sim 8.2\times 10^{-5}~M_\odot$ for the
outbursts P1 and P2, respectively.

At the end of the outbursts P1 and P2 \emph{i.e.}  when the source undergoes a
transition from the high/soft state to the low/hard state, the count rate
drops very quickly over a timescale less than a month (see
Fig.~\ref{fig_LC}). As shown by Servillat et al. (2011), any disk component in
the spectrum of HLX-1 in the low/hard state is marginal.  Given the XMM3
luminosity of $\sim 2\times 10^{40}$ erg s$^{-1}$ derived from the powerlaw
component and assuming a radiation efficiency in the low/hard state of 0.11 as
during the outbursts, we could compute an upper limit on the accretion rate in
the low/hard state of $\dot{m} < 0.09$ \emph{i.e.}  $\dot{M} < 3.6\times
10^{-6}~M_\odot$ yr$^{-1}$. This is nicely consistent with the results
obtained by Esin et al. (1997, 1998) using the Advection Dominated Accretion
Flow model. According to that model, this must be connected with a varying
(receding) inner disc radius (e.g. Dubus et al. 2001). Higher
statistical quality of the data when the source transits to the low/hard state
will be needed to further investigate that possibility.

\subsection{The $L-T$ relation}
\label{relation}

In the luminosity-temperature space, we found a correlation $L\propto
T^{\sim2.4}$.  Fukue (2000) and Kawaguchi (2003) showed that near the
Eddington limit ($\dot{m}=10$) and above a departure from the $L - T^4$
relation is expected, due to the change of the radial profile of the disk
temperature and the $\dot{m}$-sensitive disk colour temperature via the
effects of electron scattering.  However below this limit, the disk is still
expected to follow the $L\propto T^4$ relation found in the standard disk
model. Apart from the peak of the outbursts when the accretion rate is at the
Eddington limit, the accretion rate is below this limit. Therefore, we should
expect to have a $L\propto T^4$ correlation. The behaviour near the peak is
probably responsible for the flatter correlation. To check that, we fitted our
$L-kT$ points for the $XMM2$, $S_{CR1}$, $S_{CR2}$, $S_{CR3}$, $S_{t3}$
observations (for which $\dot{m} < 10$ - see Table~\ref{tab_slimdisk}) using
the relation for a standard disk. We found: $\log(L) = 4~\log~T +44.95\pm
0.06$ ($1\,\sigma$ error). From Eq.~1 in Lasota et al. (2011) giving $L$ as a
function of the colour disk temperature at the inner disk radius and the BH
mass, we derived the following relation: $\log(L) \sim 4~\log~T + 45.1$ using
$\eta \sim 0.11$, $M=1.8\times 10^4$~M$_\odot$ and $R_{ISCO} = 3\,R_S$ (for a
non-rotating BH). This demonstrates that the disk in HLX-1 follows the
$L\propto T^4$ relation.

\section{Conclusion}

We investigated in detail the X-ray spectral properties of the best IMBH
candidate, HLX-1, using multi-epoch data collected by three different X-ray
instruments ({\it Swift}-XRT, {\it XMM-Newton} EPIC-pn and {\it Chandra}). To
do so, we used the K03 disk model in order to constrain the BH mass, the
accretion rate and the disk structure in a non biased way. Indeed, this disk
model has the advantage of taking into account the effects of radial advection
and covering a wide range of accretion rates (from sub- to super-Eddington)
assuming a non-rotating BH and a face-on disk inclination ($i=0^\circ$). We
found that our multi-epoch data are consistent with sub-Eddington accretion
onto a nearly $2\times 10^4$ M$_\odot$ BH. At the peak, the X-ray luminosity
is near the Eddington luminosity. The derived radiation efficiency is
$\eta\sim 11\%$.  The disk solution we found for HLX-1 is different from those
derived for other ULXs using the same disk model (e.g. Vierdayanti et
al. 2006, Okajima et al. 2006, Yoshida et al. 2010). Indeed, for other ULXs
the spectral fitting favours super- or near Eddington accretion onto a stellar
mass BH. Here, the disk in HLX-1 likely undergoes sub- and near Eddington
accretion onto an IMBH and appears to stay relatively thin. The inner regions
of the accretion disk are dominated by radiation-pressure. We also showed that
the source globally follows the $L\propto T^4$ relation. At the outburst peak
and for a further few days, the BH accretes at near the Eddington limit
($\dot{m}\sim 10$ i.e. $\sim 4\times 10^{-4}$ M$_\odot$ yr$^{-1}$). The
occurence of this ``plateau'' at the outburst peaks could provide some
evidence that the outburst mechanism is driven by enhanced mass transfer rate
onto a pre-existing accretion disk as proposed by Lasota et al. (2011). After
the ``plateau'', the accretion rate decreases exponentially until the end of
the outburst. However, it is not clear from our data what is the geometry of
the accretion flow when the source transits to the low/hard state, even if the
presence of an optically thin advection dominated flow would avoid a dramatic
variation in the accretion rate by more than two orders of magnitude from the
peak to the low/hard state. Finally, the recent outburst starting in August
2011 gives more credit to the $\sim 1$ year recurrence timescale proposed by
Lasota et al. (2011). Our dedicated multi-wavelength (radio, optical and
X-ray) observations of this outburst will shed further light on the nature of
this unique source as well as on the outburst mechanism.

\section{Acknowledgments}

We thank the anonymous referee for his useful comments that helped to
improve the paper.  We are grateful to Ken Ebisawa and Takashi Okajima for
creating a new fits model for {\scriptsize XSPEC} incorporating the extended
black hole mass ranges. We thank Shane Davis for his useful suggestions during
the writing of this paper. MS acknowledges supports from NASA/Chandra grants
GO0-11063X, DD0-11050X and NSF grant AST-0909073. SAF acknowledges funding
from the UK Science and Technology Funding Council. SAF is the recipient of an
Australian Research Council Postdoctoral Fellowship, funded by grant
DP110102889. VB acknowledge support from the UK STFC research council. JPL
acknowledges support from the French Space Agency CNES.

\appendix

\section{Appendix material}

\subsection{Impact of the K03 disk model computing options on
  the shape of the emergent spectra, the BH mass and the
  accretion rate}

\label{invest}

The K03 model makes use of different options to compute the
emergent spectra starting from a local BB and then including different
physical effects (advection, electron opacity, relativistic effects and
Comptonization). First, we investigate the impact of these different options
on the shape of the emergent spectra as well as on the $M$ and $\dot{M}$
parameters. We take the example of the $S_{1p+2p}$ spectrum as an
illustration, but what is described below is also valid for the other spectra.

\begin{figure}
\includegraphics[angle=0,scale=0.5]{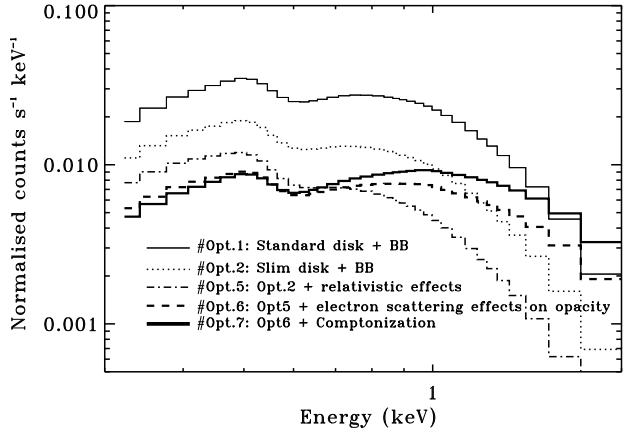}\\
\caption{Evolution of the spectral shape of the Kawaguchi (2003) disk model
  when adding different physical components to an initial standard disk model
  (red). The latter model was obtained fitting the $S_{1p+2p}$ spectrum.}
\label{fig_test}
\end{figure}

Fig.~\ref{fig_test} shows the modification of the $S_{1p+2p}$ spectrum when
selecting different model options: 1) standard disk \& local BB; 2) slim disk
\& local BB; 5) slim disk including relativistic effects on the BB; 6) slim
disk including the effects of electron scattering on opacity and relativistic
effects; 7) slim disk including the effects of electron scattering on opacity,
Comptonization and relativistic effects. Here, the term ``slim disk'' implies
that the K03 code automatically computes (at each radius) how
much advection is present for a given accretion rate.

First, we just changed the option parameters without running a new fit for
option 1 in order to see the effects of the different model components. When
using option 1, we got for the $S_{1p+2p}$ spectrum a BH mass around
$1.8\times 10^3~M_{\odot}$ and $\dot{m} \sim 32$ assuming a viscosity
parameter of $\alpha =0.01$. Such high $\dot{m}$-values mean super-Eddington
accretion. However, in this case, radial advection is expected to start
playing a significant role.  The large derived $\dot{M}$-value also implies
that electron opacity and Comptonisation should play an important role (see
Fig.2 in Kawaguchi 2003 for instance). This results in more spectral boosting
towards high energies.  The spectrum in red in Fig.~\ref{fig_test} shows the
global shape of the data. In order to get such a shape when including all the
above effects in a slim disk (option 7), it is clear that the effects of both
electron scattering on opacity and Comptonisation have to be reduced. To do
so, the accretion rate (in the unit of the Eddington rate) has to decrease;
which in turn results in an increase of the BH mass.

To check this, we fitted the $S_{1p+2p}$ spectrum for each chosen option
leaving free the model parameters. The evolution of the model parameters are
summarised in Table~\ref{tab_test}. From option 1 to option 2, there is a
relatively small variation of the $M$ and $\dot{m}$ parameters ($M\sim
3.4\times 10^3~M_{\odot}$ and $\dot{m} \sim 38$).  From option 2 to option 5
(including the relativistic effects), we obtained a significant impact on the
accretion rate ($\dot{m} \sim 200$) and a slight decrease of the BH mass
($M\sim 2.3\times 10^3~M_{\odot}$).  Given the large accretion rate
($\dot{m} \sim 38$) derived in option 2, the accretion disk extends below the
last stable orbit (e.g. Watarai et al. 2000). The relativistic effects
(gravitational redshift and transverse Doppler redshift) when applying option
5 strongly suppress the emission coming from these regions; which in turn
results in the model underestimating the data. To have a good match between
the data and the model, the accretion rate has to be increased.  From option
5 to option 6, the spectral parameters significantly change with an increase
in the BH mass by one order of magnitude and a decrease of the accretion rate
by more than one order of magnitude. From option 6 to option 7, the parameters
do not change significantly ($M \sim 1.8 \times 10^4$~M$_\odot$ and $\dot{m}
\sim 10$).

From option 5, the accretion flow is in the super-Eddington regime. The
effects of Comptonisation and electron scattering (option 6) are likely to be
significant given the large $\dot{m}$ value (see Fig.~2 in K03). This would in
turn result in a stronger hardening of the model than what is seen in
Fig.~\ref{fig_test}. However, in order to obtain a good fit given the shape of
the $S_{1p+2p}$ spectrum, the effects of both Comptonisation and electron
scattering has to be strongly minimised. This is achieved by significantly
decreasing the accretion rate (the accretion flow is then in the sub-Eddington
regime) and therefore by significantly increasing the BH mass. If we force the
source distance from 95 Mpc (measured distance) to an arbitrary lower value
(e.g. $d=3.5$ Mpc), such a dramatic change in the BH mass and the accretion
rate is not seen because the accretion flow always stays in the sub-Eddington
regime (see Table~\ref{tab_test2}).

\subsection{Comparison of the K03, KERRBB \& BHSPEC results in the limit of the Novikov \&
  Thorne (1973) relativistic disk model}

\label{invest2}

We compare the results from the K03 disk model with those obtained with
KERRBB and BHSPEC assuming a non-spinning BH ($a^*=0$) and a face-on disk
($i=0^\circ$) in the limit of low mass accretion rate (i.e. less than $10\%$
of the Eddington limit).  In that limit, the accretion is expected to follow
the predictions of the relativistic disk model from Novikov \& Thorne
(1973). In the K03 model, this limit corresponds to $\dot{m}\le
1.6~L_{Edd}/c^2$. For the comparisons with KERRBB, we limit ourselves by
setting the hardening factor to 1 in the model (\emph{i.e.} we did not take
the effects of electron scattering and Comptonisation into account). We also did
not consider the effects of limb-darkening and self-irradiation since
they are not included in the K03 disk model. This configuration corresponds to
option 5 in the K03 disk model (only including the relativistic effects). For
the comparisons with BHSPEC, we used the option 7 (including the effects for
Comptonisation and electron scattering) in the K03 disk model. We used the
table of BHSPEC models
bhspec$_{-}$mass$_{-}$0.01.fits\,\footnote{http://www.cita.utoronto.ca/$\sim$swd/xspec.html}
considering BH mass up to $300~M_\odot$.

To perform the comparisons, we proceeded in both cases as follows:
\begin{itemize}

\item we fitted the $S_{1p+2p}$ spectrum (i.e. when the source luminosity is
maximum) using the WABS*KERRBB and WABS*BHSPEC model with $a^*=0$, $i=0^\circ$
and $N_H = 4\times 10^{20}$ cm$^{-2}$ for different and arbitrary values of
the source distance $D$ so that the $\dot{m}$ values derived were below $10\%$ of
the Eddington limit.

\item from the best-fits, we simulated a model spectrum with an exposure time
   of 38 ks using the {\it Swift}-XRT response files and the fakeit command in
   Xspec.

\item we fitted the faked spectra with the WABS*K03 model with alpha fixed to
  0.01, the same source distance $D$ as used for the KERRBB and BHSPEC fits and
  the option 5 and 7 when comparing the results with KERRBB and BHSPEC,
  respectively.
\end{itemize}

Tables~\ref{tab1} and~\ref{tab2} summarise the results of the comparisons with
KERRBB and BHSPEC, respectively. From Table~\ref{tab1}, we show that there is
a systematic discrepancy (less than 45\% in the range considered) in the BH
mass between the two models, while the accretion rate values appear to be
consistent within the errors. From Table~\ref{tab2}, we obtained a good
agreement between the BHSPEC and K03 models within the $3\,\sigma$ errors
except for the BH mass estimate for a distance of 2.9 Mpc. Based on
that latter case, the ``systematic errors'' in the BH mass are less than 24\%.
We also show in Tables~\ref{tab1} and~\ref{tab2} comparisons with $\dot{m}$
values larger than $10\%$ of the Eddington limit. Again, we emphasize the good
agreement between the K03-KERRBB models and K03-BHSPEC models. So, we conclude
that the K03 disk model provides reliable estimates of accretion disk
parameters (mass and accretion rate).


\begin{table*}
\begin{center}
\caption[]{Summary of the results obtained for both the KERRBB and
  K03 models. The errors are given at the 90\% confidence level.}
\label{tab1}
\begin{tabular}{lllll}
\hline
Distance  & Parameter & KERRBB$^\circ$ &  K03$^\triangle$ & $\dot{m}^\dagger$\\
(Mpc)          &           &        &      & ($L_{Edd}/c^2$)\\  
\hline
$10.$ & Mass ($M_\odot$) & $145$ & $191^{+42}_{-29}$ $^*$ & \\

$10.$ & $\dot{m}$~($10^{19}$ g s$^{-1}$) & $21.6$ & $18.8^{+5.1}_{-3.8}$ & ($6.8^{+1.1}_{-0.9}$)\\
\hline
5. & Mass ($M_\odot$) & $73$ & $105^{+7}_{-8}$ &\\
5. & $\dot{m}$~($10^{19}$ g s$^{-1}$) & $5.4$ & $5.3^{+0.4}_{-0.4}$ & ($3.5^{+0.1}_{-0.1}$)\\
\hline
2.9 & Mass ($M_\odot$) & $42$ & $52^{+9}_{-7}$ &\\
2.9 & $\dot{m}$~($10^{19}$ g s$^{-1}$) & $1.8$ & $1.5^{+0.4}_{-0.3}$ & ($2.0^{+0.3}_{-0.2}$)\\
\hline
2. & Mass ($M_\odot$) & $29$ & $39^{+4}_{-3}$ &\\
2. & $\dot{m}$~($10^{19}$ g s$^{-1}$) & $0.85$ & $0.75\pm 0.08^*$ & ($1.32^{+0.07}_{-0.07}$)\\
\hline
1.5 & Mass ($M_\odot$) & 22 & $32^{+2}_{-5}$ &\\
1.5 & $\dot{m}$~($10^{19}$ g s$^{-1}$) & 0.49 & $0.50^{+0.04}_{-0.08}$ & ($1.08^{+0.04}_{-0.03}$)\\

\hline
\end{tabular}
    \begin{list}{}{}
    \item $^\circ$ The effects of Comptonisation, electron scattering,
    limb-darkening and self-irradiation were not taking into account.
    \item $^\triangle$ We used the option 5 in the K03 model that only includes the
    relativistic effects. 
    \item $^\dagger$ Accretion rate value derived from the K03 disk model.
    \item $^*$ Consistent at the $3~\sigma$ level.
    \end{list}
\end{center}
\end{table*}

\begin{table*}
\begin{center}
\caption[]{Summary of the results obtained for both the BHSPEC and
  K03 models. The errors are given at the 90\% confidence level.}
\label{tab2}
\begin{tabular}{lllll}
\hline
Distance  & Parameter & BHSPEC &  K03$^\triangle$ & $\dot{m}^\circ$\\
(Mpc) &  &  &   & $L_{Edd}/c^2$\\
\hline
7. & Mass ($M_\odot$) & $234$ & $237^{+73}_{-21}$ &\\
7. & $l = L_X/L_{Edd}$ & $0.146$ & $0.200^{+0.06}_{-0.04}$ $^*$ & ($3.20^{+0.06}_{-0.04}$)\\
\hline
5. & Mass ($M_\odot$) & $158$ & $172^{+31}_{-23}$ &\\
5. & $l$ & $0.110$ & $0.116^{+0.024}_{-0.020}$ & ($1.9^{+0.2}_{-0.2}$)\\
\hline
2.9 & Mass ($M_\odot$) & $85$ & $105^{+8}_{-8}$ $^\dagger$ &\\
2.9 & $l$ & $0.069$ & $0.073^{+0.006}_{-0.006}$ & ($1.18^{+0.04}_{-0.04}$)\\
\hline
2. & Mass ($M_\odot$) & $54$ & $55^{+6}_{-5}$ &\\
2. & $l$ & $0.051$ & $0.064^{+0.007}_{-0.006}$ & ($1.02^{+0.03}_{-0.02}$ $^\ddagger$)\\

\hline
\end{tabular}
    \begin{list}{}{}
    \item $^\triangle$ We used the option 7 in the K03 model that includes the
    relativistic effects as well as the effects of Comptonisation and electron
    scattering. 
    \item $^\circ$ Accretion rate value derived from the K03 disk model.
    \item $^*$ Consistent at the $3~\sigma$ level. 
    \item $^\dagger$ Marginally consistent at $3~\sigma$.
    \item $^\ddagger$ Lower error bar pegged to the lowest permitted value of
    the model ($\dot{m}=1$). 
    \end{list}
\end{center}
\end{table*}

\begin{deluxetable}{cccccc}
\tablecolumns{7}
\tablewidth{0pc}
\tablecaption{Log of the {\it XMM-Newton}, {\it Chandra} and {\it Swift}
  observations. \label{tab_obs}}
\tablehead{
\colhead{Obs. name} & Instrument & \colhead{Obs. ID} & \colhead{Start date} & \colhead{End date}  &
\colhead{Good time}\\
 &  & & (yyyy-mm-dd) &  (yyyy-mm-dd) & \colhead{(ks)}}
\startdata
XMM1 & {\it XMM-Newton} EPIC-pn & 0204540201 & 2004-11-23 & 2004-11-23 &  22 \\
XMM2 & {\it XMM-Newton} EPIC-pn, & 0560180901 & 2008-11-28 & 2008-11-28 &  51 \\
XMM2     &   {\it XMM-Newton}  RGS1, RGS2  &  0560180901 &  2008-11-28 & 2008-11-28 & 51\\

XMM3 & {\it XMM-Newton} EPIC-pn & 0655510201 & 2010-05-14 & 2010-05-15 &  62  \\
\hline
Chandra & {\it Chandra} ACIS & 13122 & 2010-09-06 & 2010-09-07&  10 \\
\hline
- & {\it Swift}-XRT & 00031287(001-139) & 2008-10-24 & 2011-10-10 & 360 \\
\enddata
\end{deluxetable}

\begin{deluxetable}{ll}
\tablecolumns{7}
\tablewidth{0pc}
\tablecaption{Nomenclature of the {\it Swift}-XRT spectra. \label{tab_nomenclature}}
\tablehead{
\colhead{Spectrum name} & Comment \\}
\startdata
$S_{1p}$ & time when the source reached its maximum luminosity \\
 & for the first outburst P1 (a count rate of $CR > 0.02$ cts s$^{-1}$) \\

$S_{2p}$ & same but for the second outburst P2 ($CR > 0.02$ cts s$^{-1}$)\\
 
$S_{1p+2p}$ & Sum of the $S_{1p}$ and $S_{2p}$ spectra \\

$S_{after~2p}$ & $\sim 7$ days after the peak of the second outburst P2 ($CR > 0.02$ cts s$^{-1}$)\\ 

\hline

$S_{CR0}$ & $0.01< CR <0.02$ cts s$^{-1}$\\

$S_{CR1}$ & $0.015 < CR < 0.02$ cts s$^{-1}$\\ 

$S_{CR2}$ & $0.01 < CR < 0.015$ cts s$^{-1}$\\

$S_{CR3}$ & $0.004 < CR < 0.01$ cts s$^{-1}$\\

$S_{CR4}$ & $0.002 < CR < 0.004$ cts s$^{-1}$\\ 

$S_{CR5}$ & $CR < 0.002$ cts s$^{-1}$\\

\hline

$P3^-$ & Outburst in August 2011 ($55788 <~\mathrm{MJD}~< 55809$) \\
$P3^+$ & Outburst in August 2011 ($55710 <~\mathrm{MJD}~< 55844$) \\
\hline
$S_{t1}$ & $55788 <~\mathrm{MJD}~< 55802$ \\
$S_{t2}$ & $55803 <~\mathrm{MJD}~< 55822$ \\
$S_{t3}$ & Reflare event in the P3 outburst ($55826 <~\mathrm{MJD}~< 55844$) \\

\enddata
\end{deluxetable}

\begin{deluxetable}{llcccccccc}
\tablecolumns{10} \tablewidth{0pc} \tablecaption{Summary of the spectral
parameters when fitting the {\it Swift}-XRT data using the {\scriptsize
DISKBB} model. \label{tab_diskbb}}

\tablehead{

\colhead{(1)} & \colhead{(2)} & \colhead{(3)} & \colhead{(4)} & \colhead{(5)}
& \colhead{(6)} & \colhead{(7)} & \colhead{(8)} & \colhead{(9)} & \colhead{(10)}\\

\colhead{Name$^*$} & \colhead{$kT$} & \colhead{$N_{BB}$} &
\colhead{$\Gamma$} & $N_\Gamma$ & $HR$ & \colhead{$L_{tot}$} &
\colhead{$L_{disk}$} & \colhead{$\log(L_{bol})$} & \colhead{$\chi^2/dof$}}
 
\startdata

$S_{1p+2p}$ &  $0.24\pm0.01$ & $20^{+6}_{-5}$  &
... & ... & $0.06\pm0.01$ & $12.0\pm0.4$ & $12.0\pm0.4$ & $42.2\pm0.03$ & 42/36\\
\hline

$S_{1p}$ & $0.24\pm0.02$ & $20^{+8}_{-6}$  &
... & ... & $0.06\pm0.01$ & $12.0\pm0.5$ & $12.0\pm0.5$ & $42.19^{+0.04}_{-0.05}$ & 18/21\\

$S_{2p}$ & $0.22\pm0.02$ & $30^{+14}_{-10}$  &
... & ... & $0.04^{+0.01}_{-0.02}$ & $13.3\pm0.7$ & $13.3\pm0.7$ & $42.23^{+0.05}_{-0.06}$ & 24/15\\

\hline
$S_{after~2p}$ & $0.18\pm0.02$ & $71^{+58}_{-33}$ & ... &...&
$0.022^{+0.006}_{-0.008}$  & $13.3\pm 0.8$ &  $13.3\pm 0.8$ & $42.25\pm 0.07$ & 54/67\\

$S_{CR0}$ & $0.18^{+0.02}_{-0.01}$ & $36^{+14}_{-12}$ & $2.2^{+0.5}_{-0.6}$ & $2.9^{+1.8}_{-1.4}$ & $0.20^{+0.04}_{-0.06}$ &
$7.5^{+0.3}_{-0.6}$ & $5.6^{+0.3}_{-0.6}$ & $41.90\pm0.04$ & 54/67\\

$S_{CR3}$ & $0.17\pm0.05$ & $11^{+36}_{-8}$ & $2.2^{+0.5}_{-0.6}$  &
$1.8\pm 1.5$ & $0.37\pm0.15$ & $3.3^{+0.4}_{-0.8}$ & $2.0^{+0.3}_{-0.7}$ & $41.46^{+0.10}_{-0.13}$ & 54/67\\

$S_{CR4}$ & ... & ... & $2.2^{+0.5}_{-0.6}$ & $2.0\pm0.7$ &
$1.2^{+0.9}_{-0.5}$ & $1.3\pm 0.2$ & ... & ...& 54/67\\

$S_{CR5}$ &  ... & ... &  $2.2^{+0.5}_{-0.6}$ & $0.3^{+0.3}_{-0.2}$ & $1.2\pm 0.9$ & $0.20\pm
0.06$ & ... & ... & 54/67\\

\hline

$S_{CR1}$ & $0.18\pm 0.02$ & $41^{+25}_{-19}$ & $2.2^\dagger$ & $2.9\pm 1.2$ &
$0.15^{+0.08}_{-0.06}$ & $9.7^{+0.9}_{-1.6}$ & $7.8^{+0.8}_{-1.5}$ & $42.03\pm0.06$ & 45/51\\

$S_{CR2}$ & $0.17\pm 0.02$ & $40^{+27}_{-17}$ &  $2.2^\dagger$ & $3.0^\pm 0.7$ & $0.22^{+0.06}_{-0.07}$ & $6.7^{+0.7}_{-0.8}$
& $4.8^{+0.3}_{-0.5}$ & $41.85\pm0.06$ & 45/51\\

\hline

$S_{t1}$ &  $0.20\pm 0.02$ & $39^{+25}_{-15}$ & ... & ... &
$0.029^{+0.012}_{-0.010}$ & $12.4\pm 2.0$ & $12.4\pm 2.0$ & $42.22\pm0.07$ & 15.2/15\\

$S_{t2}$ &  $0.20\pm 0.02$ & $41^{+16}_{-12}$ & ... & ... &
$0.023^{+0.006}_{-0.008}$ & $10.5\pm 1.0$ & $10.5\pm 1.0$ & $42.15\pm0.05$ & 17.2/22\\

$S_{t3}$ &  $0.18\pm 0.03$ & $44^{+57}_{-25}$ & $2.2^\dagger$ & $2.8^{+1.9}_{-2.0}$ &
$0.18^{+0.13}_{-0.08}$ & $8.6\pm1.7$ & $6.8\pm 1.2$ & $41.98\pm0.10$ & 9.5/7\\

\enddata
    \begin{list}{}{}
    \item Columns - (1) spectrum name; (2) inner disk temperature in units of
    keV ; (3) the black-body normalisation; (4) Photon index of the powerlaw
    component; (5) Normalisation of the powerlaw in units of $10^{-5}$ photons
    keV$^{-1}$ cm$^{-2}$ s$^{-1}$; (6) hardness ratio defined as the ratio of
    the observed 1.5-10 keV flux over the observed 0.3-10 keV flux; (7)
    unabsorbed 0.2-10 keV total luminosity in units of $10^{41}$ erg s$^{-1}$;
    (8) unabsorbed 0.2-10 keV disk luminosity in units of $10^{41}$ erg
    s$^{-1}$; (9) Logarithm of the unabsorbed 0.01-20 keV disk luminosity in
    units of $10^{41}$ erg s$^{-1}$; (10) $\chi^2$ value and number of degrees
    of freedom.

    \item $^*$ See Table~\ref{tab_nomenclature} for the nomenclature of the
    different spectra.

    \item $^\dagger$ Fixed parameter.
    \end{list}

\end{deluxetable}

\begin{deluxetable}{rrr}
\tablecolumns{3} \tablewidth{0pc} \tablecaption{Summary of the $p$-value
derived using the {\scriptsize DISKPBB} model for the {\it XMM-Newton}, {\it
Swift}-XRT and {\it Chandra} data. \label{tab_diskpbb}}

\tablehead{
\colhead{Spectrum} & \colhead{$p$} }
\startdata
\colhead{XMM1} & \colhead{$0 < 0.54$}\\
\colhead{XMM2} & \colhead{$0.63^{+0.37}_{-0.11}$}  \\
\hline
\colhead{Chandra$^\dagger$}   & \colhead{$0.57^{+0.33}_{-0.07}$} \\
\hline
\colhead{$S_{1p}$}   & \colhead{$0.71\pm 0.2$} \\
\colhead{$S_{2p}$}   & \colhead{$0.53-1$} \\
\colhead{$S_{after~2p}$}   & \colhead{$0.55^{+0.45}_{-0.05}$} \\
\colhead{$S_{CR0}$}   & \colhead{$0.7^{+0.3}_{-0.2}$} \\

\enddata
    \begin{list}{}{}
    \item $^\dagger$ The value was obtained fixing $N_H$ to $4 \times 10^{20}$
    cm$^{-2}$. 
    \end{list}
\end{deluxetable}

\begin{deluxetable}{lccccccccc}
\tablecolumns{10} \tablewidth{0pc} \tablecaption{Summary of the spectral
parameters when fitting the {\it XMM-Newton}, {\it Swift} and {\it Chandra}
data using the Kawaguchi (2003) disk model. \label{tab_slimdisk}}

\tablehead{

\colhead{(1)} & \colhead{(2)} & \colhead{(3)} & \colhead{(4)} & \colhead{(5)}
& \colhead{(6)} & \colhead{(7)} & \colhead{(8)} & \colhead{(9)} & \colhead{(10)}\\

\colhead{Spectrum} & \colhead{$N_H$} & \colhead{$M$} & \colhead{$\dot{M}$} &
\colhead{$\alpha$} & $\Gamma$ & $N_\Gamma$ & \colhead{$L_{tot}$} & \colhead{$L_{disk}$} & \colhead{$\chi^2/dof^\dagger$}}
 
\startdata
XMM1 & 4 & $1.4\pm 0.1$ & $4.8\pm0.3$ & $0.11^{+0.08}_{-0.05}$ & ... & ... &
$4.2^{+0.2}_{-0.6}$ & $4.2^{+0.2}_{-0.6}$ & 283/215 \\
XMM2 & 4 &  $1.4\pm 0.1$ & $5.9\pm0.4$ & $0.010^{+0.002}_{-0.000}$ & $1.6^{+0.3}_{-0.4}$ & $1.0\pm0.4$  &
$5.1^{+0.1}_{-0.3}$ & $4.3^{+0.1}_{-0.2}$ & 283/215 \\

XMM1 & $5.5^{+1.1}_{-1.0}$ & $1.9^{+1.3}_{-0.3}$ & $4.4^{+0.3}_{-0.5}$ &
$0.13\pm 0.07$ & ... & ... & $4.9^{+0.2}_{-0.6}$ & $4.9^{+0.2}_{-0.6}$& 277/214 \\
XMM2 & $5.5^{+1.1}_{-1.0}$ &  $1.9^{+1.3}_{-0.3}$ & $5.4^{+0.5}_{-0.7}$ & $0.010^{+0.005}_{-0.000}$ & $1.8^{+0.4}_{-0.5}$ &
$1.3^{+0.4}_{-0.6}$ & $5.7^{+0.1}_{-0.5}$ & $4.9^{+0.1}_{-0.3}$& 277/214 \\

\hline
Chandra & 4 & $1.9\pm0.2$ & $10.2^{+0.8}_{-2.1}$ & $0.025^{+0.009}_{-0.015}$ &... & ... &
$10.7^{+0.3}_{-0.4}$ & $10.7^{+0.3}_{-0.4}$ & 38/28\\
Chandra & $3.7^{+3.9}_{-2.5}$ & $1.8^{+1.0}_{-0.4}$ & $10.2^{+0.7}_{-2.7}$ &
$0.023^{+0.017}_{-0.013}$ &... & ... & $10.5^{+0.5}_{-0.6}$ & $10.5^{+0.5}_{-0.6}$ & 37.6/27\\
\hline
$S_{1p+2p}$ & 4 & $1.8^{+0.2}_{-0.1}$ & $10.1^{+0.3}_{-2.3}$  &
$0.05^{+0.04}_{-0.02}$ & ... & ... & $11.0^{+0.4}_{-0.7}$ & $11.0^{+0.4}_{-0.7}$ & 108/104\\
$S_{after~2p}$ &4 & $1.8^{+0.2}_{-0.1}$ & $10.5^{+0.3}_{-1.1}$ & $0.015^{+0.005}_{-0.005}$ &...&... & $9.9^{+4.2}_{-0.3}$ &$9.9^{+4.2}_{-0.3}$ & 108/104\\
$S_{CR0}$ & 4& $1.8^{+0.2}_{-0.1}$ & $5.5\pm0.5$ & $0.020^{+0.009}_{-0.006}$ & $1.7^{+0.6}_{-0.8}$ & $1.4^{+1.6}_{-0.9}$ &
$6.2^{+0.3}_{-0.7}$ & $5.1^{+0.2}_{-0.4}$ & 108/104\\
$S_{CR3}$ & 4& $1.8^{+0.2}_{-0.1}$ & $3.8\pm 0.4$ & 0.01 fixed & $1.7^{+0.6}_{-0.8}$ & $1.6^{+1.5}_{-1.0}$ & $3.3^{+0.6}_{-0.7}$ & $2.3\pm0.5$ & 108/104\\

\hline
$S_{CR1}$ & 4 & 1.8 fixed & $7.1\pm0.9$ & $0.016^{+0.028}_{-0.006}$ & $1.6^{+0.9}_{-1.3}$ &
$1.3^{+2.5}_{-1.0}$ & $8.0^{+0.5}_{-0.9}$ & $6.9^{+0.4}_{-0.8}$ & 46/50\\

$S_{CR2}$ & 4 & 1.8 fixed & $4.9^{+0.3}_{-0.5}$ & $0.019^{+0.024}_{-0.009}$ & $1.6^{+0.9}_{-1.3}$ & $1.3^{+2.5}_{-1.1}$ & $5.5^{+0.4}_{-0.6}$
& $4.4^{+0.4}_{-0.5}$ & 46/50\\

\hline

$S_{t1}$ & 4 & $1.8$ fixed & $10.9^{+0.3}_{-1.5}$ & $0.01^{+0.11}_{-0.00}$ & ... & ... & $10.8\pm1.1$
& $10.8\pm1.1$ & 37.7/41\\

$S_{t2}$ & 4 & $1.8$ fixed & $9.1\pm 1.0$ & $0.016^{+0.005}_{-0.006}$ & ... & ... & $8.8\pm0.6$
& $8.8\pm0.6$ & 37.7/41\\

$S_{t3}$ & 4 & $1.8$ fixed & $6.7^{+1.4}_{-1.2}$ & $0.018^{+0.023}_{-0.007}$ & ... & ... & $6.7\pm0.9$
& $6.7\pm0.9$ & 37.7/41\\

\enddata
    \begin{list}{}{}
    \item Columns - (1) spectrum name; (2) absorption column in units of
    $10^{20}$ cm$^{-2}$; (3) BH mass in units of $10^4~M_\odot$; (4) accretion
    rate in units of $\frac{L_{Edd}}{c^2}$ with $L_{Edd}$ and $c$, the
    Eddington limit and the speed of light, respectively; (5) viscosity
    parameter; (6) Photon index of the powerlaw component; (7) Normalisation
    of the powerlaw in units of $10^{-5}$ photons keV$^{-1}$ cm$^{-2}$
    s$^{-1}$; (8) unabsorbed 0.3-10 keV total luminosity in units of $10^{41}$
    erg s$^{-1}$; (9) unabsorbed 0.3-10 keV disk luminosity in units of
    $10^{41}$ erg s$^{-1}$; (10) $\chi^2$ value and number of degrees of
    freedom.
    \item $^\dagger$ The {\it Swift}-XRT, {\it XMM-Newton} EPIC-pn and {\it
    Chandra}-ACIS data were fitted separately. For a given instrument, we fitted
    together all the available spectra or a sub-set of them.  
    \end{list}

\end{deluxetable}


\begin{deluxetable}{ccccc}
\tablecolumns{5} \tablewidth{0pc} \tablecaption{ Evolution of the parameters of the Kawaguchi (2003) slim disk model for different model
options. The fits were performed using the $S_{1p+2p}$ spectrum with $N_H=4\times 10^{20}$ cm$^{-2}$.  \label{tab_test}}

\tablehead{
\colhead{Option} & \colhead{$M$} & \colhead{$\dot{m}$} &
\colhead{$\alpha$} & \colhead{$chi^2/dof$} \\
\colhead{(1)} & \colhead{(2)} & \colhead{(3)} & \colhead{(4)} & \colhead{(5)}}
\startdata
1 & $1.8^{+0.4}_{-0.2}$ & $32.2^{+1.1}_{-5.6}$ & $0.01$ fixed & 40.4/36 \\
2 & $3.5^{+0.3}_{-0.2}$ & $38.0^{+1.4}_{-1.3}$ & $0.82^{+0.18}_{-0.55}$ & 40.1/35 \\
5 & $2.3^{+1.0}_{-0.6}$ & $200^{+477}_{-101}$ & $>0.26$ & 51.59/35 \\
6 & $29^{+4}_{-14}$ & $10.0^{+1.7}_{-0.9}$ & $0.14^{+0.01}_{-0.04}$ & 67.21/35 \\
7 & $18\pm 2$ & $10.1^{+0.3}_{-2.3}$ & $0.05^{+0.06}_{-0.02}$ & 43.4/35 \\
\enddata
    \begin{list}{}{}
\item Columns - (1) Computation option from the Kawaguchi (2003) slim disk
    model (see Section~\ref{invest}); (2) BH mass in units of $10^3~M_\odot$;
    (3) Accretion rate in units of $\frac{L_{Edd}}{c^2}$; (4) viscosity
    parameter; (5) value of the $\chi^2$ and number of degree of freedom.
    \end{list}
\end{deluxetable}

\begin{deluxetable}{ccccc}
\tablecolumns{5} \tablewidth{0pc} \tablecaption{Evolution of the parameters of
the Kawaguchi (2003) slim disk model for different model options forcing the
source distance to $3.5$ Mpc instead of the measured distance of 95 Mpc.
\label{tab_test2}}

\tablehead{
\colhead{Option} & \colhead{$M$} & \colhead{$\dot{m}$} &
\colhead{$\alpha$} & \colhead{$chi^2/dof$} \\
\colhead{(1)} & \colhead{(2)} & \colhead{(3)} & \colhead{(4)} & \colhead{(5)}}
\startdata
1 & $79^{+15}_{-12}$ & $1.07^{+0.05}_{-0.04}$ & $0.01$ fixed & 40.2/36 \\
2 & $64^{+15}_{-10}$ & $1.7^{+0.2}_{-0.2}$ & $0.01$ fixed & 40.3/36 \\
5 & $58^{+14}_{-8}$ & $2.9^{+0.4}_{-0.5}$ & $0.01$ fixed & 40.8/36 \\
6 & $52^{+5}_{-5}$ & $3.3^{+0.1}_{-0.1}$ & $0.01$ fixed & 40.4/36 \\
7 & $122^{+13}_{-10}$ & $1.4^{+0.1}_{-0.1}$ & $0.01$ fixed & 42/36 \\
\enddata
    \begin{list}{}{}
\item Columns - (1) Computation option from the Kawaguchi (2003) slim disk
    model (see Section~\ref{invest}); (2) BH mass in units of $M_\odot$;
    (3) Accretion rate in units of $\frac{L_{Edd}}{c^2}$; (4) viscosity
    parameter; (5) value of the $\chi^2$ and number of degree of freedom.
    \end{list}
\end{deluxetable}

\end{document}